\documentclass[sn-mathphys]{sn-jnl}

\jyear{2021}%

\usepackage{lineno,hyperref}
\usepackage{changepage}
\usepackage{algorithm}
\usepackage{multirow}
\usepackage{epsfig}
\usepackage{subfig}
\usepackage{enumitem}
\usepackage{array}
\usepackage{footnote}

\algnewcommand{\Inputs}[1]{%
  \Statex \textbf{Inputs:}
  \Statex \hspace*{\algorithmicindent}\parbox[t]{.8\linewidth}{\raggedright #1}
}
\algnewcommand{\Outputs}[1]{%
  \Statex \textbf{Outputs:}
  \Statex \hspace*{\algorithmicindent}\parbox[t]{.8\linewidth}{\raggedright #1}
}
\algnewcommand{\Initialize}[1]{%
  \State \textbf{Initialize:}
  \Statex \hspace*{\algorithmicindent}\parbox[t]{.8\linewidth}{\raggedright #1}
}

\theoremstyle{thmstyleone}%
\theoremstyle{thmstyletwo}%

\theoremstyle{thmstylethree}%

\newcolumntype{M}[1]{>{\centering\arraybackslash}m{#1}}
\newcolumntype{P}[1]{>{\centering\arraybackslash}p{#1}}

\begin{document}

\title[TCP using test case diversification and fault-proneness estimations]{Test case prioritization using test case diversification and fault-proneness estimations}


\author[1]{\fnm{Mostafa} \sur{Mahdieh}}

\author*[1]{\fnm{Seyed-Hassan} \sur{Mirian-Hosseinabadi}}\email{hmirian@sharif.edu}

\author[1]{\fnm{Mohsen} \sur{Mahdieh}}

\affil*[1]{\orgdiv{Computer Engineering Department}, \orgname{Sharif University of Technology}, \orgaddress{\street{Azadi Ave}, \city{Tehran}, \country{Iran}}}


\abstract{Regression testing activities greatly reduce the risk of faulty software release. However, the size of the test suites grows throughout the development process, resulting in time-consuming execution of the test suite and delayed feedback to the software development team. This has urged the need for approaches such as test case prioritization (TCP) and test-suite reduction to reach better results in case of limited resources. In this regard, proposing approaches that use auxiliary sources of data such as bug history can be interesting. We aim to propose an approach for TCP that takes into account test case coverage data, bug history, and test case diversification. To evaluate this approach we study its performance on real-world open-source projects. The bug history is used to estimate the fault-proneness of source code areas. The diversification of test cases is preserved by incorporating fault-proneness on a clustering-based approach scheme. The proposed methods are evaluated on datasets collected from the development history of five real-world projects including 357 versions in total. The experiments show that the proposed methods are superior to coverage-based TCP methods. The proposed approach shows that improvement of coverage-based and fault-proneness-based methods is possible by using a combination of diversification and fault-proneness incorporation.}




\maketitle

\section{Introduction}
\label{sec:introduction}
Modern software systems are continuously changing at a rapid rate throughout the software development process. Changes are made to the software to add new features, improve functionality, and repair identified software bugs. During this evolution, developers mustn't unintentionally inject new bugs into the software. Software \emph{regression testing} attempts to reduce this risk by running a certain suite of test cases regularly or after each modification. Due to the increasing size of the software codebase and the number of change commits, regression testing has become a resource-intensive procedure for current software projects. Moreover, re-running the test suite can take much time, which results in a large feedback delay for developers. For example, at Google, code modification commits are done at the rate of more than 20 times per minute requiring more than 150 million test executions per day \cite{kumar2010development, memon2017taming}.

There has been a wide range of techniques proposed to improve the cost-effectiveness of regression testing. These techniques can be categorized into three groups: Test suite reduction, Test case selection, and Test case prioritization. \textit{Test suite reduction} (also referred to as Test suite minimization) techniques speed up regression testing by reducing the size of the test suite~\cite{zhong2008experimental, fraser2007redundancy, chen2017assertions}. These methods try to eliminate repetitive test cases, in hope of creating a smaller test suite with similar fault detection capability. \textit{Test case selection} techniques intend to prevent unnecessary regression testing by choosing test cases that cover the modified code between versions~\cite{grindal2006evaluation, yoo2007pareto, kazmi2017effective}. \textit{Test case prioritization (TCP)} techniques aim to reorder test cases such that early fault detection is maximized~\cite{jones2003test, elbaum2002test}. This approach has the advantage over test suite reduction and test case selection in that it does not exclude any test cases from execution. TCP methods provide a way to execute test cases with more fault detection ability earlier to provide early feedback to developers. TCP also allows continuing testing to the limit of time or budget, by running the test suite in order obtained by prioritization. 

Test case prioritization, which is the subject problem of this study, has been highly investigated and many approaches have been proposed for the TCP problem. \cite{lou2019survey, khatibsyarbini2018test, hemmati2019advances}. The majority of TCP methods have used structural code coverage as a metric to prioritize test cases \cite{hao2014unified, yoo2012regression}. Some researchers have investigated using other sources of information, such as the project requirements \cite{hettiarachchi2016risk, srikanth2016requirements, salehie2011prioritizing}, source code changes, \cite{alves2016prioritizing, saha2015information, panda2016slice} or test execution history \cite{noor2015similarity, khalilian2012improved, rahman2018prioritizing}.

One valuable source of information for TCP is the bug history of the project. Bug history has been recently proposed as a source of information to improve TCP \cite{wang2017qtep, paterson2019empirical, eghbali2019supervised, mahdieh2020incorporating}. Bug history can be utilized to estimate the \textit{fault-proneness} of code units, which is the probability that developers have injected a defect in a code unit. In this line, defect prediction methods have been employed to estimate the fault-proneness of code units based on the source code and bug history of the project \cite{zimmermann2007predicting, ostrand2005predicting, menzies2006data}. However, there remains a challenge in the strategy of incorporating the fault-proneness estimations obtained by defect prediction to prioritize the test cases. For example, if fault-proneness is naively used to prioritize test cases, the test cases that cover a fault-prone area, although similar and possibly redundant will have a high priority.

It has been intuitively conjectured that test cases that have similar properties, also are probable to have similar fault detection capability \cite{leon2003comparison, yoo2009clustering}. Therefore considering diversification in selecting and prioritizing test cases will lead to appropriate results. Based on this conjecture, various approaches have been proposed for TCP \cite{jiang2009adaptive, fang2014similarity, ledru2012prioritizing}. These methods have been shown to empirically improve the fault detection rate of TCP, confirming the mentioned conjecture about similar test cases' fault detection capability in the context of TCP.


In this paper, we propose a novel approach to incorporate fault-proneness estimations for TCP utilizing both fault-proneness and test case diversification. Our approach is based on the idea of grouping similar test cases using clustering methods and prioritizing the clustered test cases. To estimate the fault-proneness of all code units from the bug history and source code, we designed a defect prediction method customized for the regression testing setting. Furthermore, we developed a TCP method based on test case clustering which takes into account both fault-proneness and test case diversification.

Another challenge regarding TCP is the empirical study. Many studies are based on manually injected faults or mutant versions of programs. To measure the fault detection rate of various TCP strategies in a more realistic situation, we evaluate the algorithms on real-world projects containing defects that occurred in the development process. Our experiment is conducted on 357 versions of five real-world projects, included in the Defects4J dataset \cite{just2014defects4j}, comparing the fault detection rate of multiple methods.


To more accurately assess our study, we raise the following research questions:
\begin{itemize}
    \item {\bf RQ1:}  The traditional total and additional TCP strategies have proven to be successful for coverage-based TCP \cite{rothermel2002empirical, hao2015optimal, hao2014unified}. How do the proposed clustering-based TCP methods compare to the traditional coverage-based TCP strategies in terms of fault detection performance?
    \item {\bf RQ2:} Does incorporating fault-proneness improve the proposed clustering-based TCP algorithm in terms of fault detection performance?
    \item {\bf RQ3:} What is the influence of the clustering parameters (distance function and the number of clusters) on the effectiveness of the proposed TCP algorithms?
\end{itemize}
\vspace{\baselineskip}
This paper makes the following contributions:
\begin{itemize}
\item We provide an approach to leverage existing coverage-based TCP methods in a clustering-based TCP scheme. This approach led to the development of new TCP methods which are based on test case coverage data and take advantage of the diversification of test cases for TCP.
\item We propose a novel approach to combine fault history data and test case diversification, in the context of coverage-based TCP.
\item We design a customized defect prediction method to estimate the fault-proneness of a code unit. This method is customized to work when only a small set of recorded bugs are available and utilizes the information from all versions of the source code history.
\item
We present an empirical evaluation using five open-source projects containing in total 357 versions of the projects. Results show that our proposed approach could improve existing coverage-based TCP techniques.
    
\end{itemize}

The rest of the paper is organized as follows: Section~\ref{sec:background} presents the background material. Section~\ref{sec:methodology} presents our approach to solving the problem and our proposed method. Section~\ref{sec:empirical_study} presents the setup of our empirical evaluation and Section~\ref{sec:results} shows the results of our experiments. In Section~\ref{sec:discussion} the empirical results and threats to the validity of this study are discussed. Section~\ref{sec:related_work} summarizes the most related work to this paper. Finally, Section~\ref{sec:conclusions} contains the conclusions and future work of this paper.

\section{Background}
\label{sec:background}
In this section, we present the formal definition of test case prioritization and briefly introduce some of the classical coverage-based TCP methods. We continue by providing background information on defect prediction, which is employed for fault-proneness estimation in this study. Afterward, we present concepts of test case similarity and diversification-based methods for TCP.

\subsection{Test case prioritization}
In its essence, TCP seeks to find a permutation of test cases, which optimizes a certain intended goal function. To more formally define TCP, consider a test suite containing the set of test cases $T = \{ t_1, t_2, \ldots, t_n \}$. The TCP problem is defined as follows \cite{elbaum2002test}:

\begin{adjustwidth}{1cm}{}
{\it Given:} $T$, a test suite; $PT$, the set of permutations of T; $f$, a function from $PT$ to the real numbers.\\	
{\it Problem:} Find $T' \in PT$ such that\footnote{This relation is expressed using Z notation's first order logic \cite{usingz}.}:
\begin{equation}
\label{prioritization_problem_1} 
\forall T'' : PT \mid T'' \ne T' \bullet f(T') \ge f(T'')
\end{equation}

\end{adjustwidth}

In other words, the TCP problem is finding a permutation $T'$ such that $f(T')$ is maximized. Here $f$ is a scoring function that assigns a score value to any permutation selected from $PT$.

The $f$ function represents the goal of a TCP activity. Software engineers using TCP methods could have different goals, such as testing business-critical functionality as soon as possible, maximizing code coverage, or detecting faults at a faster rate. Since the ultimate target of regression testing is to detect regression faults, the TCP target function is usually specified as to how fast the regression faults can be detected, which is referred to as \textit{fault detection rate}. One of the highly used measurements for evaluating the fault detection rate is the APFD (Average Percentage of Faults Detected) goal function, an area-under-curve metric that measures how quickly a test suite can detect faults. APFD is frequently used in the literature for TCP when the goal of TCP is maximizing the fault detection rate \cite{yoo2012regression, engstrom2011improving, catal2013test}. Another target function that can be used is the percentage of test cases executed until the first failing test case. We have chosen the first-fail metric for our empirical study and discussed this in~\ref{sec:subjects_of_study}.


\subsection{Coverage-based test case prioritization}
\label{coverage_based_tcp}
For the sake of modeling the system for test case prioritization, the source code can be partitioned into units such as files, methods, or statements. Assuming a chosen level of partitioning, the source is partitioned into units $U = \{u_1, u_2, \ldots, u_m \}$. Using this modeling, a broad range of coverage-based TCP methods settle on a level of partitioning (usually statements or methods) and measure coverage of test cases over those units. Considering each test case $t_i$ of the test suite and unit $u_j$ of the source code, $Cover(i, j)$ represents how much test case $t_i$ covers unit $u_j$. The amount of coverage can be either 0 or 1 if the units of code are statements; however if the units are methods or files, it can also be a real number in the range $[0,1]$ representing the proportion of code that is covered by the test case execution.

Test case coverage can be collected in different ways. Dynamic coverage information is collected by executing the test case and tracking every unit of code that is executed. On the other hand, static coverage is derived by static analysis of the source code \cite{mei2012static}. 

When the coverage of a test case on the code units is known, other concepts such as the total coverage of the test case can be computed. The total coverage of a test case $t_i$ is formally defined as follows\cite{hao2014unified}:
\begin{equation}
\label{total_coverage_equation}
Cover(i) = \sum\limits_{1 \le j \le m} Cover(i,j)
\end{equation}

The value of $Cover(i)$ is a real non-negative number and can be larger than 1. Coverage-based TCP methods utilize coverage of test cases to prioritize the test suite. Traditional coverage-based TCP methods will be reviewed in the following subsection.

\subsection{Review of traditional TCP methods}
\label{sec:traditional_strategies}
In this subsection, we review three traditional TCP strategies that are considered baseline methods in our empirical study.

\subsubsection{Random strategy}
The obvious and simple method of TCP is the random strategy. Taking the random strategy, all test cases of the test suite are shuffled in random order. The expected first-fail metric and APFD of this strategy are near $50\%$. This method is usually presented as the first baseline to be compared with other proposed strategies for evaluation \cite{ashraf2012value, elbaum2004selecting}.

\subsubsection{Total strategy for TCP}
\label{sec:total_prioritization}
The \textit{total prioritization strategy} is based on the intuition that test cases that have more coverage are more likely to uncover bugs. The total strategy, therefore, starts with computing the total coverage of all test cases according to ~Equation~\ref{total_coverage_equation}. In the next step, test cases are sorted according to their total coverage and as the result, the first test case in the prioritized order has the highest total coverage. 
The total prioritization strategy does not consider the fact that some test cases might cover duplicate areas of the code. Therefore, when test cases are prioritized using this strategy, frequently some units of code are executed multiple times before the whole units are covered \cite{elbaum2002test}. 

Compared to other non-random existing strategies, the total prioritization strategy is simple and efficient. The time complexity of this algorithm is the sum of the time complexity of computing the total coverage for all test cases and the time complexity of the sorting algorithm. The addition of these values results in the time complexity of the total algorithm which is $\mathcal{O}(nm + n\log{}n)$, where $n$ is the number of test cases and $m$ is the number of source code units.

\subsubsection{Additional strategy for TCP}
\label{trad_add_strat}

In contrast to total prioritization, the \textit{additional prioritization strategy} takes into account that executing an uncovered unit of the code is more likely to reveal new faults in the code, and therefore a test case that runs uncovered code must have more priority compared to a test case that runs already covered units. The idea behind the additional strategy is that earlier coverage of uncovered units of the code, results in revealing faults sooner \cite{elbaum2002test}.

The additional strategy begins by computing the total coverage of all test cases. Afterward, in each step, the test case with the highest coverage over the uncovered code area is chosen as the next test case. The selected test case is appended to the end of the list of prioritized test cases and marked not to be chosen in the next steps. The area of the code covered by the selected test case will be marked as a covered area. 

With this type of selection, the additional strategy falls in the category of greedy algorithms. This strategy works in $n$ steps where $n$ is the number of test cases. In each step, selecting the next test case and updating the coverage of the remaining test cases is done in time complexity of $\mathcal{O}(nm)$. Therefore, the total time complexity of this algorithm is $\mathcal{O}(n^2 m)$.

Due to different implementations in some scenarios, different variations of The additional strategy have been developed. In two situations, this strategy faces different options:
\begin{itemize}
\item When there exist at least two non-selected test cases which both have the highest coverage over the uncovered code area. In case of such a tie, one of these test cases should be selected with some criteria. For example, one might select the test case randomly.
\item In case there are no uncovered areas of the code left. In this case, the remaining test cases can be ordered with different approaches. A common solution is to consider all the code uncovered again and continue the algorithm with the remaining test cases \cite{elbaum2002test}. 
\end{itemize}

\subsection{Defect Prediction}
\label{sec:background_defect_prediction}
Software faults are an inevitable part of the development process. These faults happen for various reasons such as the addition or modification of the software features, lack of tests and documentation, high level of dependence between units, and faulty designs. 

Modern software development tools can track and record occurrences of each fault. As the cause behind most code faults is related to a limited set of known or unknown generic fault patterns, it is reasonable to generalize the pattern using the previously recorded samples.



There are usually four major steps to a defect prediction method~\cite{nam2014survey}:
\begin{enumerate}
\item{Feature extraction}: In this step, each unit of code (package, file, class, or method) is analyzed and various metrics are extracted from the unit. The result of this step is a feature vector for each unit plus a label that indicates whether the unit contains bugs or not.
\item{Data preprocessing}: To maximize the quality of defect prediction algorithms, the extracted data should be manipulated in accordance with the machine learning algorithm. This step includes removing unnecessary features, normalization, and sampling.
\item{Model learning}: A machine learning algorithm is selected to predict faulty code based on previous versions. The extracted feature vectors are then fed to the machine learning algorithm. A small portion of the training samples is reserved for validation. The choice of the algorithm is made based on the quality of the predictions made by the model on the validation set. Prediction quality is then evaluated by the model's performance on the test set. 
\item{Prediction}:
The last step is to predict defects in unseen samples. In this step, each new unit of code is labeled with a fault-proneness score, indicating the plausibility of a defect in the unit.
\end{enumerate}

There have been various features proposed for defect prediction. Static code metrics which mainly capture the complexity and structural aspects of the source code have been proposed, such as McCabe \cite{mccabe1976complexity}, Halstead metrics \cite{halstead1977elements}, CK
features (design metrics from UML) \cite{chidamber1994metrics}, and object-oriented features (coupling, cohesion,
etc.) \cite{harrison1998evaluation, bansiya2002hierarchical, e1994candidate}.
Many studies have used static code metrics for defect prediction \cite{menzies2010defect,menzies2007data,zimmermann2007predicting}. Other metrics, such as historical and process-related metrics (e.g., number of past bugs \cite{klas2010transparent,ostrand2005predicting} or the number of changes \cite{pinzger2008can, meneely2008predicting, moser2008comparative}) and organizational metrics (e.g., number of developers \cite{weyuker2008too, graves2000predicting}), have also been proposed.


Various machine learning techniques have been explored for the prediction step of defect prediction, which can be categorized as supervised learning, unsupervised learning, and semi-supervised learning~\cite{li2018progress}. Many supervised classification models have been applied for defect prediction, such as decision trees~\cite{menzies2006data}, neural networks~\cite{kanmani2007object}, support vector machines~\cite{elish2008predicting}, Naive Bayes~\cite{shivaji2009reducing}, and Bayesian networks~\cite{okutan2014software}. Jing et al.~\cite{jing2014dictionary} employed cost-sensitive dictionary learning for defect prediction. More recently, ensemble learning methods have shown interesting performance and have gained attention in the area of software defect prediction~\cite{petric2016building, aljamaan2020software, matloob2021software, li2019software}. Unsupervised learning has been employed for defect prediction~\cite{li2020systematic} based on clustering methods \cite{nam2015clami, bishnu2011software, zhang2016cross} and other unsupervised approaches~\cite{yang2016effort, fu2017revisiting, yan2017file, boucher2018software}. Semi-supervised learning methods have been utilized for defect prediction using sparse learning~\cite{wang2016non} and graph-based label propagation~\cite{zhang2017label}.
The prediction of bugs at change-level or commit-level, namely Just-In-Time (JIT) software defect prediction was introduced by Kamei et al.~\cite{kamei2012large}. 

In recent years deep learning techniques have also been utilized for defect prediction. Yang et al. \cite{yang2015deep} utilize a deep belief network to extract a set of expressive metrics from an initial set of change metrics. Using the extracted features their method trains a classifier to predict defects at the change-level. Wang et al. \cite{wang2016automatically, wang2018deep} leveraged deep belief networks to learn semantic features from abstract syntax trees and then used these features to create defect prediction models. 
Hoang et al.~\cite{hoang2019deepjit} introduced \textit{DeepJIT} for Just-In-Time defect prediction, which utilizes Convolutional Neural Network (CNN) in an end-to-end deep learning framework by extracting features from both commit messages and code changes. 
Hoang et al.~\cite{hoang2020cc2vec} proposed \textit{CC2Vec} as an improvement to DeepJIT using a Hierarchical Attention Network (HAN) architecture. Pandey et al.~\cite{pandey2020bpdet} present \textit{BPDET} for defect prediction, by implementing a two-layer ensemble of different classifiers in front of an autoencoder-based deep representation.
Popular deep learning methods that have been applied in defect prediction, include Long Short-Term Memory (LSTM) \cite{majd2020sldeep, deng2020software, liang2019seml}, Stacked Denoising Autoencoder \cite{tong2018software, zhu2020within}, CNN \cite{li2017software}, and Deep Neural Network (DNN) \cite{xu2019ldfr}. 
Despite promising results using deep learning methods, applying presents new challenges which are under investigation. Yedida et al.~\cite{yedida2021value}, point out that many researchers applying deep learning in software engineering tasks have not compared the results with other non-deep learning techniques. They also provide experiments showing that class imbalance issues still should be cared for when using deep learning methods for defect prediction. In their study they show that using deep learning methods without applying appropriate preprocessing techniques such as oversampling might significantly decrease effectiveness of these methods.

\subsection{TCP based on fault-proneness estimations}
\label{sec:tcp_based_on_fp}
In case there is prior knowledge available on the presence of faults in certain areas of the code, this knowledge can be employed to improve test case prioritization. One of the categories of research in this line is based on estimating fault-proneness using defect prediction methods \cite{wang2017qtep, paterson2019empirical, eghbali2019supervised, mahdieh2020incorporating}. In \cite{mahdieh2020incorporating}, fault-proneness based coverage is presented, which is defined as:
\begin{equation}
\label{fp_based_cover}
Cover^{FP}(i) = \sum\limits_{1 \le j \le m} Cover(i,j) \times Prob(F_j)
\end{equation}
Where $Cover^{FP}(i)$ denotes the fault-proneness based coverage of test case $t_i$ of the test suite and the estimated probability\footnote{$F_j$ indicates the event in which \textit{j}th code unit is faulty and $Prob(F_J)$ represents the probability of this event.} of existing faults in unit $u_j$ ($1 \le j \le m$) is shown by $Prob(F_j)$.

This concept of coverage can be incorporated in coverage-based TCP methods such as the ones presented in Section~\ref{sec:traditional_strategies}.

\subsection{Diversity based TCP}
\label{sec:diversity_based_tcp}
As mentioned in Section~\ref{sec:introduction}, it is believed that test cases with similar features have similar fault detection capabilities \cite{leon2003comparison, yoo2009clustering}. The general idea behind the diversity-based TCP approach is to rank the test cases in an order that at each point of execution of the test cases, the diversity of the executed test cases at that point is maximized.
To do so, these methods attempt to implement the following three steps: 
\begin{itemize}
\item Encode test cases as a vector of features
\item Computing distance/similarity of test cases according to a distance metric
\item Maximize/minimize the distances/similarities of test cases
\end{itemize}

There have been different distance functions proposed for diversity-based TCP such as Euclidean distance, Hamming distance, Jaccard Index, and Edit distance \cite{hemmati2019advances}. 

After choosing an appropriate distance function, an algorithm must be determined to order test cases such that the diversity of the test cases along the prioritized test cases sequence is maximized. The problem of prioritizing test cases to achieve such
maximum diversity is an NP-hard problem (traditional set cover) \cite{mathur2010foundations}. Therefore a heuristic method must be used to attempt to find a permutation of the test cases which sub-optimally maximizes the diversity function. 
Various heuristic methods have been proposed for this maximization problem. These methods can be categorized as Greedy, Adaptive Random, Clustering, and Search-based algorithms \cite{hemmati2019advances}.

Among these categories, clustering-based methods have been employed by several researchers for TCP \cite{kandil2017cluster, pei131dynamic, fu2017coverage, shrivathsan2019novel, chen2018test}. Clustering methods partition data points into groups or clusters, according to the similarity function between the data points, such that data points in the same group have high similarity. For the TCP application, normally a data point is extracted from each test case, which represents the properties of the test case. Various clustering algorithms can be applied in this scenario, such as hierarchical clustering, centroid-based clustering, clustering based on fuzzy theory, distribution-based clustering, density-based clustering, and clustering based on graph theory \cite{xu2015comprehensive}. After clustering the data points various strategies can be imagined to prioritize the test cases. We review some of the previously proposed strategies in Section~\ref{sec:related_work}. This paper also leverages clustering and proposes a strategy in this manner.

\section{Methodology}
\label{sec:methodology}
\begin{figure}[b]
\centering
\includegraphics[width=4in]{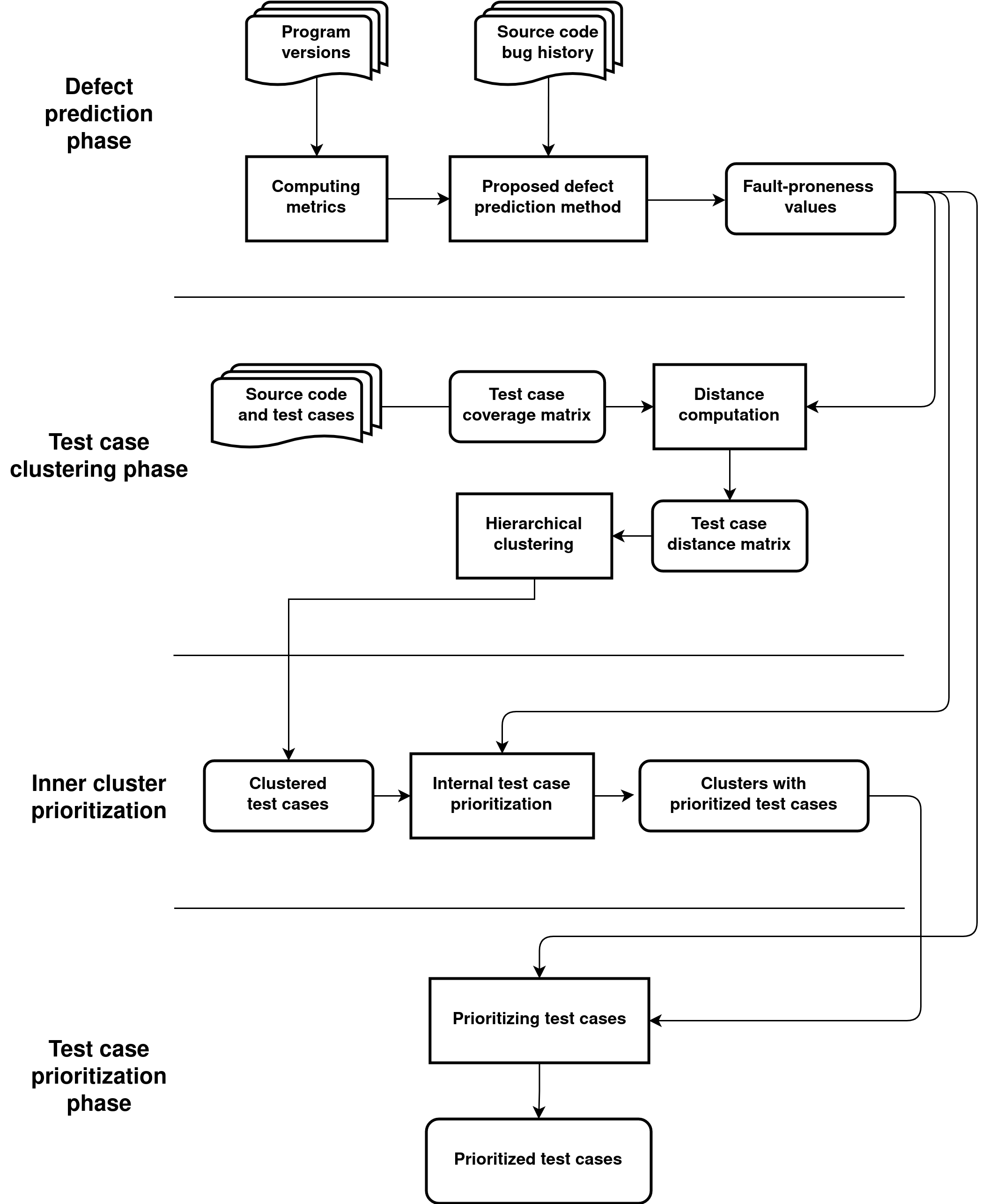}
\caption{Overview of the proposed test case prioritization algorithm}
\label{fig:proposed_tcp_algorithm}
\end{figure}

Our proposed method consists of four main steps. The block diagram of the proposed method is depicted in Figure~\ref{fig:proposed_tcp_algorithm}. In the first phase, defect prediction is utilized to predict the fault-proneness of code units. In the second phase, the test cases are clustered to similar test cases in groups using a hierarchical clustering algorithm. In the third phase, the test cases in each cluster are internally prioritized using coverage-based TCP methods. In the fourth phase test cases of the clusters are aggregated combining the fault-proneness estimations and traditional TCP methods. 
To evaluate the prioritization algorithm, the actual test results are used to find the rank of the fault-revealing test cases and to compute the fault detection rate of each of the algorithms. 

Two proposed algorithms are derived from this approach:
\begin{itemize}
\item The first step can be skipped resulting in the proposed TCP method not utilizing fault-proneness and defect prediction concepts. Algorithm~\ref{alg:proposed_tcp_nofp} shows the pseudo-code of the TCP method without using fault-proneness. We refer to this algorithm as \texttt{CovClustering} in the empirical study (Section~\ref{sec:empirical_study}).
\item Algorithm~\ref{alg:proposed_tcp_with_fp} specifies the pseudo-code of the method with the incorporation of fault-proneness derived by defect prediction. We refer to this algorithm as \texttt{CovClustering+FP} in the empirical study results.
\end{itemize}

In the following subsections, we explain each of the main steps of the proposed approach in more detail. Explanations of the pseudo-codes follow in Section~\ref{sec:proposed_tcp_final_steps}, where we also give more details on the third and fourth steps of the algorithms.

\subsection{Proposed defect prediction method}
\label{sec:prop_def_pred_method}
In this section, the problem of predicting defective codes and the proposed method to obtain an appropriate estimate of the fault-proneness are presented. As mentioned in Section~\ref{sec:tcp_based_on_fp}, the goal is to predict $Prob(F_i)$ for each unit $u_i$ of code given a set of features extracted from this unit. One effective method to estimate $Prob(F_i)$ is modeling it through a binary classification model. 

In this section, we describe the concepts of the defect prediction model, and the  details of implementing this model are described in Section~\ref{sec:defect_prediction_implementation}.

\subsubsection{Feature set}
The project data used for this study, including code features and code coverage data are described in detail in Section~\ref{sec:subjects_of_study}. The features are extracted at the class level and consist of 104 input features including static and process features shown in Table~\ref{tab:features}. Each recorded version of the project in the dataset contains exactly a single bug that can be detected by a few test cases of the test suite. The classes which are buggy will result in data instances marked with the buggy label and the other instances are marked as not-buggy instances.

One helpful step to achieving better performance is to combine features to get new meaningful features. For each version, if there are no changes to a specific unit and other units related to it, it would be highly unlikely that it turns into a buggy unit and vice-versa. 
Using this idea, a set of features is defined by calculating the difference between two consecutive version metrics of each unit.
This set of features can also be seen as the first derivative in discrete time by viewing the unit versions as the time dimension. 
This additional set of features (known as the churn of source code metrics) is shown to be effective for defect prediction~\cite{d2012evaluating}, therefore we add them to the feature set.
Keeping the original features and the churn of source code features together helps to capture both the static and dynamic aspects of the units.

We reviewed defect-prediction methods based on deep learning in Section~\ref{sec:background_defect_prediction}, however, as mentioned applying these techniques still has difficulties and can be challenging.
Our goal is to focus on designing a TCP method, which should successfully work with any good enough defect-prediction method. On the other hand, the volume of source code contained in the subject  of empirical study also presents limitations for employing deep learning methods. Therefore we consent to consider more simple defect-prediction models in this paper if their performance is acceptable. Note that using more simple methods has the benefits of less implementation complexity and lower execution resources needed which can be interesting.

\subsubsection{Classification Model}
As mentioned in Section~\ref{sec:background_defect_prediction}, classification is one of the main components of the defect prediction procedure. Various models have been utilized to be used to predict bugs \cite{nam2014survey}. Lessmman et al.~\cite{lessmann2008benchmarking} compare many classification models for defect prediction in a standard benchmark and the results show that the classification model has a negligible effect on the performance of defect prediction. Among 22 classification models compared in this study, Random Forest has the best performance which confirms the result of previous studies~\cite{guo2004robust}.  Ghotra et al.~\cite{ghotra2015revisiting} revisit these studies by applying noise-cleaning to the NASA dataset and adding the PROMISE dataset, and observe that tree-based ensemble methods have the best performance. Aljamaan et al.~\cite{aljamaan2020software} study tree-based ensemble models for defect-prediction and confirm that these methods have promising performance and specifically observe that random forest and XGBoost have notable results. These models are a good fit for our scenario of defect prediction as they have the following properties:

\begin{itemize} 
    \item  
        Our dataset consists of numerous features both originating directly from the codes or derived in the data preparation phase. Although removing irrelevant features using feature selection methods might help with the tree building process, selecting the most effective features manually is a difficult task. An important advantage of tree-based methods is that these models inherently select the most effective features based on the training labels (target values) implicitly. 
    \item
        Using tree-based models could be helpful since the result only considers the samples which are in the leaf node. Therefore it almost resistant to most of the preprocessing issues which must be considered. However, it's still useful to remove the noisy samples and generate more samples of the class with a smaller size. 
    \item
        These models are invariant to most scaling methods and require little to no normalization. 
    \item Ensemble models are less prone to over-fitting. 
\end{itemize}

These properties lead to the effectiveness of tree-based ensemble defect prediction models on generally any software project and decrease the dependency of the results on a specific dataset~\cite{hastie2008elements}. In section~\ref{sec:defect_prediction_implementation} we present the results of empirically comparing multiple tree-based defect-prediction models on the subject dataset, which leads to the conclusion that XGBoost has the best performance on the subject dataset of study. Therefore XGBoost was chosen as the classification model in the proposed defect prediction method.

An important step is to tune the key parameters of the XGBoost model. The parameter tuning process is usually defined as an optimization problem and the goal is to optimize a certain scoring function. We selected this approach and used it iteratively and the parameters were chosen using validation data. The details of this approach are mentioned in Section~\ref{sec:defect_prediction_implementation}.

\subsubsection{Data preprocessing}
\label{sec:data_preprocessing}

Machine learning models can benefit from data standardization before feeding the data to the model. To standardize the data, it's important to understand the requirements of the selected ML model. In tree-based models (Random Forest, XGBoost, etc.), the final tree structure is invariant to scaling input vectors linearly~\cite{hastie2008elements}. This is since each split is applied at a given interval. Also, the predicted target value is independent of the input vector because it is calculated using training phase target values. 

The dataset consists of over one hundred features and after adding the different features, the number of features jumps to two hundred. Having this many features makes the classification difficult. To address this problem several measures are set to place. 

The first strategy is to use the regularization parameters in the classification model. Specifically, L1 regularization is a helpful way to tackle over-fitting when dealing with sparse datasets~\cite{hastie2008elements}. The XGBoost model has implemented three regularization parameters: \begin{enumerate}
\item Alpha is the L1 regularization coefficient.
\item Lambda is the L2 regularization coefficient.
\item Gamma is the minimum loss reduction of a leaf node to be partitioned. 
\end{enumerate}
All three of these parameters are carefully examined and taken into effect.

The second strategy is manually limiting the max depth of the trees. This is a straightforward approach since it limits the choice of dimensions used in the trees and avoids the curse of dimensionality problem. 

The third strategy is to sample the number of columns used for each tree in the boosting model. This way, when adding a new tree, the number of features making an effect on it is limited. Hence, it reduces the chance of an over-fitted model.

In real-world projects, it is very common that the project contains a few recorded bugs for each version~\cite{khoshgoftaar2010attribute}. Due to this fact, for each version, there are mostly non-faulty samples. In machine learning terms, this phenomenon is known as imbalanced classification and standard machine learning algorithms struggle in this case and must be implemented with care~\cite{song2018comprehensive}. Hence, some steps in the data preparation should be manipulated to tackle these problems.

There are several well-known approaches to deal with the imbalanced dataset issue. These include oversampling, undersampling, and class weights. Oversampling attempts to generate samples nearby the existing samples with the same label to increase the samples of the smaller class~\cite{chawla2002smote}. On the other hand, undersampling methods remove the samples that are considered noisy (i.e. two samples with different labels that are very close in the input vector space) and samples that are insignificant to the results (e.g. duplicates). 
Furthermore, the class weight approach is a really useful technique in tree-based methods since it emphasizes more on the class with the least samples. We come back to this issue and mention our approach to handling imbalanced data in Section~\ref{sec:defect_prediction_implementation}.  



\subsection{Proposed clustering method}
\label{prop_clus_pred_method}
To arrange similar test cases in groups, we use a clustering method. As explained in Section~\ref{sec:diversity_based_tcp}, a standard clustering method receives a set of points, each with a feature vector as input, and returns multiple subsets of points (i.e. the clusters) as output. In our proposed method, the points are the test cases. To create a feature vector for each test case, we use the vector of source code coverage values of that test case. This vector has a size equal to the number of methods in the source code and is created by appending the traced coverage value of each method of the source code after execution of the test case, which is a real number in the range of $[0, 1]$. The value of 0 represents no coverage of the test case on the method and the value of 1 represents coverage of all statements of the method by executing the test case.

The agglomerative hierarchical clustering method is applied to cluster the test cases. In the agglomerative clustering method, a bottom-up approach is followed. The clustering starts by considering each of the points as a cluster and follows by merging pairs of nearest clusters until the number of clusters reaches the desired number of clusters. Some studies have already reported the successful application of this clustering method in the application of test case prioritization \cite{carlson2011clustering, fu2017coverage}. One of the reasons for using a hierarchical clustering method, in this case, is that the origin of the clusters in the data and the number of clusters are not fully known. Hierarchical clustering methods do not consider any assumption on the number of clusters in the data and have more resistance to such situations.

Clustering methods use a distance function to group similar points (i.e. points with small distance) and any of the normal distance functions can be used for this purpose. Another configuration of the agglomerative clustering methods is the type of distance measurement between two clusters of points. The \textit{average} linkage metric is used in this manner. This metric is computed using the average pairwise distance between points from each cluster.

We also utilize the results of the defect prediction phase to arrange better clusters. As we are partitioning the test cases into groups, it is reasonable to arrange the partitions in a manner that all partitions have a comparable probability of fault-revealing test cases. Therefore we modify the coverage matrix such that test cases that cover areas of code with a high fault-proneness are put into different clusters. More exactly, the coverage of code units is multiplied by their corresponding $Cover^{FP}$ value (presented in Section~\ref{sec:tcp_based_on_fp}) for all coverage values and then the distance is computed based on these modified coverage values. As result, test cases covering code areas with high fault-proneness would be grouped with test cases covering the same areas and there would be resistance for them to be merged with other clusters which cover high fault-proneness areas.

\begin{algorithm}[!htb]
\caption{Proposed test case prioritization algorithm (\texttt{CovClustering})}\label{alg:proposed_tcp_nofp}
\begin{algorithmic}[1]
\Inputs{\textit{Cover:} the coverage matrix of the test suite
\\ \textit{n:} size of the test suite
\\ \textit{clusterNum:} the number of clusters for clustering}
\Outputs{\textit{Prioritized:} the prioritized list of the test suite}
\State $\mathit{D} \gets \mathsf{distances}(Cover)$
\State $\mathit{testClusters} \gets \mathsf{agglomerativeClustering}(D, \mathit{clusterNum})$
\For {each $c (1 \leq c \leq \mathit{clusterNum})$}
    \State $\mathit{testClusters[c]} \gets \mathsf{additionalPrioritization}(\mathit{testClusters[c]}, \mathit{Coverage})$
\EndFor
\State $\mathit{round} \gets 0$
\While {$\lvert \mathit{Prioritized} \rvert < n$}
    \State $\mathit{tests} \gets \emptyset$
    \For {each $c (1 \leq c \leq \mathit{clusterNum})$}
        \If {$\lvert \mathit{testClusters[c]} \rvert > \mathit{round}$}
            \State $\mathit{tests} \gets (\, \mathit{tests} \cup \mathit{testClusters[c][round]}  )\,$
        \EndIf
    \EndFor
    \State $\mathit{tests}' \gets \mathsf{totalPrioritization}(\mathit{tests}, \mathit{Coverage})$
    \State $\mathit{Prioritized} \gets \mathit{Prioritized} \mathbin\Vert \mathit{tests}'$ 
    \State $\mathit{round} \gets \mathit{round}+1$
\EndWhile
\end{algorithmic}
\end{algorithm}

\begin{algorithm}[!htb]
\caption{Proposed test case prioritization algorithm incorporating fault-proneness (\texttt{CovClustering+FP})}\label{alg:proposed_tcp_with_fp}
\begin{algorithmic}[1]
\Inputs{\textit{Metrics:} computed metrics 
\\ \textit{Model:} the learned defect prediction model 
\\ \textit{Cover:} the coverage matrix of the test suite
\\ \textit{n:} size of the test suite
\\ \textit{clusterNum:} the number of clusters for clustering}
\Outputs{\textit{Prioritized:} the prioritized order of the test suite}
\State $\mathit{{Prob}_{FP}} \gets \mathsf{defectPrediction}(\mathit{Metrics},\mathit{Model})$
\Comment{fault-proneness probability}
\State $\mathit{D} \gets \mathsf{distances}(Cover \times \mathit{{Prob}_{FP}})$
\State $\mathit{testClusters} \gets \mathsf{agglomerativeClustering}(D, \mathit{clusterNum})$
\For {each $c (1 \leq c \leq \mathit{clusterNum})$}
    \State $\mathit{testClusters[c]} \gets \mathsf{maxPrioritization}(\mathit{testClusters[c]}, Coverage, \mathit{{Prob}_{FP}})$
\EndFor

\State $\mathit{round} \gets 0$
\While {$\lvert \mathit{Prioritized} \rvert < n$}
    \State $tests \gets \emptyset$
    \For {each $c (1 \leq c \leq \mathit{clusterNum})$}
        \If {$\lvert \mathit{testClusters[c]} \rvert \geq \mathit{round}$}
            \State $\mathit{tests} \gets \mathit{tests} \cup \mathit{testClusters[c][round]}$
        \EndIf
    \EndFor
    \State $\mathit{tests}' \gets \mathsf{maxPrioritization}(\mathit{tests}, Coverage, \mathit{{Prob}_{FP}})$
    \State $\mathit{Prioritized} \gets \mathit{Prioritized} \mathbin\Vert \mathit{tests}'$ 
    \State $\mathit{round} \gets \mathit{round}+1$
\EndWhile

\end{algorithmic}
\end{algorithm}

\subsection{Proposed test case prioritization method}
\label{sec:proposed_tcp_final_steps}
As mentioned, the proposed TCP method consists of four steps. In the first step (defect prediction phase) which has been described in sections ~\ref{sec:prop_def_pred_method} and~\ref{sec:defect_prediction_implementation}, the fault-proneness of code units is estimated and this estimation is used to prioritize the test cases in the next steps. In the second step (clustering phase) which was explained in section~\ref{prop_clus_pred_method}, test cases are grouped into multiple clusters.

In this section, we will describe the third and fourth steps of the algorithm in detail. In the third step, the test cases of each cluster are prioritized internally, concerning each other. In the fourth step, we use an iterative approach. In each iteration, a test case is selected (according to the internally prioritized order) from each of the clusters. After that, the selected test cases are prioritized using a prioritization strategy. 

Algorithm~\ref{alg:proposed_tcp_nofp} shows the pseudo-code of the proposed TCP method. In lines 1-2, the distances between test cases are computed based on their coverage value and agglomerative clustering is executed to cluster the test cases. Lines 3-5, show the third step which is prioritizing the test cases internally in each cluster. For this purpose, we use the additional prioritization strategy which was mentioned in section~\ref{sec:traditional_strategies}. The additional strategy has been shown to have significant performance among coverage-based TCP strategies~\cite{hao2015optimal}.

Finally lines 6-16, describe the fourth step at which in each iteration of the while loop, a test case is selected from each cluster and added to the $\textit{tests}$ set. 
After that, the selected test case set is prioritized using the total prioritization strategy.
Using the additional prioritization is not necessary for this step, as clustering has already limited duplicate code coverage between clusters.

Algorithm~\ref{alg:proposed_tcp_with_fp} uses the same procedure as Algorithm~\ref{alg:proposed_tcp_nofp}, also adding incorporation of fault-proneness into the method. In line 1 of Algorithm~\ref{alg:proposed_tcp_with_fp}, the defect prediction method described in Section~\ref{sec:prop_def_pred_method} is executed to extract a fault-proneness value for each unit of the code. Line 2 computes the distances between test cases after element-wise multiplication of the coverage matrix into the fault-proneness vector. Lines 3-18 are the same as Algorithm~\ref{alg:proposed_tcp_with_fp}, with the difference of using $\mathsf{maxPrioritization}$ for prioritizing test sets in lines 5 and 15. We define $\mathsf{maxPrioritization}$ as sorting test cases in descending order by the maximum of fault-proneness of units covered by each test case.

\section{Empirical study}
\label{sec:empirical_study}
In this section, we explain our empirical study and discuss the results of our experiments.

\subsection{Research Questions}
\label{research_questions}
In our empirical study, we aim to answer several research questions, presented in the introduction of this paper. These research questions are stated as follows:
\begin{itemize}
    \item {\bf RQ1:} How does the proposed TCP method (without the usage of the defect prediction), compare to the traditional coverage-based TCP strategies in terms of fault detection performance?
    \item {\bf RQ2:} Does incorporating fault-proneness improve the proposed clustering-based TCP algorithm in terms of fault detection performance?
    \item {\bf RQ3:} What is the influence of the distance function and number of clusters on the effectiveness of the proposed TCP algorithms?
\end{itemize}

\subsection{Subjects of study}
\label{sec:subjects_of_study}
To evaluate TCP algorithms, the algorithms must be executed on projects with a large test suite. The test suite must reveal at least one bug for the prioritization to be meaningful. Furthermore, the source of the projects and the bug locations must be identifiable so that white box TCP methods can be applied. To apply defect prediction, the bug history of the project throughout development must also be recorded.

\begin{table}[!htb]
\captionsetup{font=normalsize}
\centering
\caption{Projects included in Defects4J initial version}
\label{tab:def}
\begin{tabular}{c|c|c|c}
\hline
	\textbf{Identifier} & \textbf{Project name} & \textbf{Bugs} & \textbf{Test classes} \\ \hline \hline

	Chart & JFreechart & 26 & 355\\ \hline
	Closure & Closure compiler & 133 & 221 \\ \hline
	Lang & Apache commons-lang & 65 & 112\\ \hline
	Math & Apache commons-math & 106 & 384\\ \hline
	Time & Joda-Time & 27 & 122\\ \hline \hline
	\textbf{Sum} & \textbf{-} & \textbf{357} & \textbf{1194}\\ \hline
\end{tabular}
\end{table}

The Defects4J collection presented by Just et al.~\cite{just2014defects4j}, fulfills the mentioned properties. In its initial published version, Defects4J provided a version history of five well-known open-source Java projects, which contain a considerable number of test cases, alongside a recorded bug history, summarized in Table~\ref{tab:def}. As these projects represent completely real-world project development, we can hope that the results can be more practically significant.

\begin{table}[!b]
\captionsetup{font=normalsize}
\centering
\caption{Defect prediction features~\cite{mahdieh2020incorporating}} \label{tab:features}
\begin{tabular}{|M{0.2cm}|M{1.2cm}|M{1.5cm}|M{2.2cm}|M{0.9cm}|M{4.2cm}|}

\hline
\textbf{\#} & \textbf{Feature type} & \textbf{Category} & \textbf{Definition} & \textbf{Count} & \textbf{General Items} \\ \hline \hline
1 & Input & Source code metrics & Used to quantify different source code characteristics & 52  & Cohesion metrics,  Complexity metrics,  Coupling metrics,  Documentation metrics,  Inheritance metrics,  Size metrics \\ \hline 
2 & Input & Clone metrics & Used to Identify the number of type-2 clones (same syntax with different variable names) & 8 & Clone Classes,  Clone Complexity,  Clone Coverage,  Clone Instances,  Clone Line Coverage,  Clone Logical Line Coverage,  Lines of Duplicated Code,  Logical Lines of Duplicated Code  \\ \hline 
3 & Input & Coding rule violations & Used for counting coding violation rules & 42  & Basic Rules,  Brace Rules,  Clone Implementation Rules,  Controversial Rules,  Design Rules,  Finalizer Rules,  Import Statement Rules,  J2EE Rules,  JUnit Rules,  Jakarta Commons Logging Rules,  Java Logging Rules,  JavaBean Rules,  Naming Rules,  Optimization Rules,  Security Code Guideline Rules,  Strict Exception Rules,  String and StringBuffer Rules,  Type Resolution Rules,  Unnecessary and Unused Code Rules  \\ \hline 
4 & Input & Git metrics & Used to count the number of committers and commits per file (these metrics could not be computed for inner classes) & 2  & Committers count, Commit counts \\ \hline 
5 & Output & Bug label & Label that shows this file is buggy in this version of the project or not & 1 & IsBuggy \\ \hline
\end{tabular} 
\end{table}

The Defects4J data set has been collected in a specific standard. For each recorded bug, Defects4J provides a faulty version of the project which contains the bug. In the faulty version, one or more failing test cases identify the bug. This helps us to locate buggy classes in each version of the source code. There is exactly one bug in each version of all projects of the Defects4J dataset, therefore after any failing test case is reached the bug is detected, and executing other failing test cases will not have significant value. Therefore we chose the first failing or first-fail metric to measure the fault detection rate, similar to other studies which have used this metric for evaluating TCP on the Defects4J dataset~\cite{paterson2019empirical, palma2018improvement, noor2015similarity, abou2021detrimental}. Furthermore, due to the small number of failing test cases, which is a single test case in some versions, the value of the first failing metric and the value of the APFD metric are near to equivalent in many versions of the dataset.

In order to obtain the coverage of the test cases and also code metrics extracted from the Defect4J source code, we have used the already created and publicly available Defects4J+M dataset\footnote{\url{https://github.com/khesoem/Defects4J-Plus-M}}\cite{mahdieh2020incorporating}. Defects4J+M is an extension of the Defects4J dataset, containing the measured test coverages and source code metrics for each version of all projects included in Defects4J.

In this dataset, dynamic coverage is used to measure coverage of test case execution on the source code. Dynamic coverage is generally more accurate than static coverage and can lead to more effective prioritization results. The coverage values in this dataset, represent the amount of coverage of each test case on each unit of the code. The coverage values are measured at the method level with a real value indicating the amount of coverage on each method.

The source code metrics of Defects4J+M are composed of a combination of static and process metrics. These metrics were computed at the class level. Table~\ref{tab:features} which is quoted from the article which introduces the Defects4J+M dataset~\cite{mahdieh2020incorporating} contains the details of each feature group contained in Defects4J+M. To use the computed metrics for defect prediction, we stored them in a vector that is used as the input feature vector by the defect prediction algorithm.


\subsection{Subject TCP algorithms}
\label{sec:subject_tcp_algorithms}
To empirically compare our proposed methods with related methods, we selected and implemented notable TCP methods. An important point to consider for selecting these TCP methods is that the information sources used by the methods must be the same as the proposed method. For example, if a TCP method uses both test coverage and software requirements as the information source for prioritization, it is not reasonable to compare this method with methods that only use test coverage for prioritization.

\begin{table}[!htb]
\captionsetup{font=normalsize}
\centering
\caption{Studied TCP algorithms}
\label{tab:tcp_algorithms_studied}
\begin{tabular}{|M{2cm}|M{2.5cm}|M{2cm}|M{4.5cm}|}

\hline
	\textbf{Algorithm} & \textbf{Identifier} & \textbf{Information sources} & \textbf{Description}\\ \hline \hline
	Total strategy & \texttt{Total} & Coverage & Total prioritization strategy described in Section~\ref{sec:traditional_strategies} \\ \hline
	Additional strategy & \texttt{Additional} & Coverage & Additional prioritization strategy described in Section~\ref{sec:traditional_strategies} \\ \hline
	Adaptive random TCP & \texttt{ART} & Coverage & Adaptive random test case prioritization proposed in~\cite{jiang2009adaptive}\\ \hline
	Proposed clustering based TCP method & \texttt{CovClustering} & Coverage & The proposed based TCP method without the usage of the phase of defect prediction, in either of the clustering or prioritization phases \\ \hline \hline
	Total strategy with fault-proneness based coverage & \texttt{Total+FP} & Coverage, Bug history and source code metrics & Total prioritization strategy using fault-proneness based test case prioritization proposed in~\cite{mahdieh2020incorporating}\\ \hline
	Additional strategy with fault-proneness based coverage & \texttt{Additional+FP} & Coverage, Bug history and source code metrics & Additional prioritization strategy using fault-proneness based test case prioritization proposed in~\cite{mahdieh2020incorporating}\\ \hline
	G-clef with greedy prioritization & \texttt{G-clef (Greedy)} & Coverage, Bug history and source code metrics & G-clef prioritization method proposed in~\cite{paterson2019empirical} \\ \hline	
	G-clef with additional prioritization & \texttt{G-clef (Additional)} & Coverage, Bug history and source code metrics & G-clef prioritization method proposed in~\cite{paterson2019empirical} \\ \hline
	Proposed TCP method with incorporating fault-proneness & \texttt{CovClustering+FP} & Coverage, Bug history and source code metrics & The proposed based TCP method which was presented in Section~\ref{sec:methodology} \\ \hline
	\end{tabular}
\end{table}

The TCP algorithms used for comparison in our empirical study our summarized in Table~\ref{tab:tcp_algorithms_studied}. These algorithms can be divided into two categories: First, are TCP algorithms that use only coverage as their information source, and second are TCP algorithms using coverage, bug history, and source code metrics as their information source. The algorithms of the first category are the following:
\begin{enumerate}
    \item The traditional total and additional prioritization methods (described in Section~\ref{sec:traditional_strategies})
    \item The adaptive random TCP algorithm proposed by Jiang et al.~\cite{jiang2009adaptive}.
    \item The proposed based TCP method of this paper without the usage of the phase of defect prediction, in either of the clustering or prioritization phases (we refer to this method as \texttt{CovClustering}).
\end{enumerate}

The algorithms of the second category, which use coverage, bug history, and source code metrics, are as follows:
\begin{enumerate}
    \item The total and additional prioritization TCP methods based on fault-proneness coverage proposed in \cite{mahdieh2020incorporating} and presented in Section~\ref{sec:tcp_based_on_fp}.
    \item The G-clef proposed by Paterson at al.~\cite{paterson2019empirical} in two variants: using either the greedy or the additional strategy as the secondary objective function.
    \item  The proposed based TCP method of this paper which was presented in Section~\ref{sec:methodology} that utilizes defect prediction (we refer to this method as \texttt{CovClustering+FP}).
\end{enumerate}

Note that another TCP algorithm that leverages fault-proneness is the QTEP method proposed by Wang et al.~\cite{wang2017qtep}. This algorithm is based on influencing fault-proneness on coverage. As the formulation of QTEP is very similar to the method presented in~\cite{mahdieh2020incorporating}, we only put the latter in the set of algorithms for comparison.

\subsection{Experimental procedure}
\label{sec:experimental_procedure}
The main part of the experiment consists of running the algorithms mentioned in Table~\ref{tab:tcp_algorithms_studied} on the projects of the Defects4J+M dataset. To create the defect prediction model for the $i$th version of the projects, the procedure explained in Section~\ref{sec:prop_def_pred_method} is performed by aggregating the data of the other projects and the data from the $1$st to $(i-1)$th versions of the same project. Since it is reasonable to create the defect prediction model using a minimum number of bugs from the same project, we created the model only for the versions of each project from some version onward. In this regard, the evaluation is done over all versions of the projects, except the oldest 5 versions of each project which are used for defect-prediction hyperparameter tuning. Additionally, other projects were added to the training set of each version to get an advantage in early versions. This addition happens to enhance the model performance even in later versions.

The source code of the methods implemented in this paper and usage instructions are put publicly available on a GitHub repository\footnote{\url{https://github.com/mostafamahdieh/ClusteringFaultPronenessTCP}}. This package contains instructions on the usage of the algorithms and replicating the results of this paper in multiple steps.

The defect prediction model is implemented using Python language and \textit{XGBoost} machine learning libraries. The clustering algorithms are implemented using the Python \textit{scikit-learn} library and the TCP algorithms are also implemented with Python language using \textit{NumPy} and \textit{pandas} libraries. The distance metric used for the agglomerative hierarchical clustering method is the euclidean distance metric, which is frequently applied when using this clustering method. The number of clusters chosen for our experiments was chosen by observing the Davies–Bouldin index, which is explained in~Section\ref{sec:rq_num_of_clusters}.


\subsection{Defect prediction implementation}
\label{sec:defect_prediction_implementation}
There are several key steps regarding the choice of the proposed classification model and its parameters. These include choosing the best classification algorithm, hyperparameter tuning, and data preparation techniques. We begin with the discussion of the classification algorithm and then the procedure upon which the hyperparameters are chosen is explained. Lastly, a few ideas that were tested regarding the imbalanced nature of data is introduced.
\subsubsection{Comparison of tree-based models}
In Section~\ref{sec:background_defect_prediction}, tree-based ensemble methods were introduced and in Section~\ref{sec:prop_def_pred_method} the classification model was further looked into. It is clear that the performance of models varies on different datasets, therefore to choose between tree-based ensemble methods, we selected three major models and compared their performance on our dataset. The selected models are Random Forest, CatBoost, and XGBoost where in this section a comparison between these tree-based models is presented.

The training and evaluation process is repeated separately on every version of each project. For each version of a project, the training set consists of data instances of older versions of that project and data instances of other projects. The idea is to maintain the model's generalization over all projects in the earlier stages and it further improves on average as new versions are added to the training set. We will refer to this type of classification as \textit{online}. The other type of execution of the classification algorithm, denoted by \textit{offline}, is to only use other projects in the training set and does not require iterating over versions. The offline execution type is only used to measure the impact of adding the previous versions into the training set.

We principally evaluate the defect prediction method performance  by Matthews Correlation Coefficient (or MCC in short). MCC is the binary version of the Pearson Correlation Coefficient that measures the similarity between the predicted labels and the true labels. This evaluation metric works well in the imbalanced dataset cases and has been suggested for usage in defect prediction applications~\cite{yao2020assessing,yao2021impact}, and also used in other fields~\cite{boughorbel2017optimal,chicco2017ten}. 

A comparison between the classification algorithms is shown in Table~\ref{tab:dp_comparison}. Interestingly, the offline runs have a good enough result without having seen any of the project instances and solely depending of data instances of other projects. This indicates that overfitting has not occurred in our classification models, because the training and evaluation instances are very different in this case.

It is observed that the online models have improvements over the offline models in most cases, which is reasonable due to feeding more training data for the online models. The best model among the six candidate models in terms of the MCC score, is the online XGBoost model, therefore we select this model for our further experiments.

\begin{table}[!htb]
\captionsetup{font=normalsize}
\centering
\caption{Comparison of different classification algorithms}
\label{tab:dp_comparison}
\begin{tabular}{c|c|c|c|c|c|c}

\hline
	\textbf{Classifier} & \textbf{Run Type} & \textbf{MCC} & \textbf{F1-score} & \textbf{Precision} & \textbf{Recall} & \textbf{AUC} \\ \hline \hline
	Random Forest & Offline & 0.71 & 0.69  & 0.75 & 0.71 & 0.992 \\ \hline
	Random Forest & Online & 0.68 & 0.67 &  0.67 & 0.71 & 0.994 \\ \hline
	Catboost & Offline & 0.87 & 0.87 & 0.88 & 0.86 & 0.992 \\ \hline
	Catboost & Online & 0.88 & 0.87 & \textbf{0.90} & 0.85 & \textbf{0.994} \\ \hline
	XGBoost & Offline & 0.87 & 0.87 & 0.88 &  0.87 & 0.992 \\ \hline
	XGBoost & Online & \textbf{0.89} & \textbf{0.89} & 0.89 & 	\textbf{0.90} & 0.992 \\ \hline
\end{tabular}
\end{table}

Table~\ref{tab:dp_performance} shows a detailed overview of the properties of the learning process of online XGBoost model on all projects. Column \textit{Evaluation versions} show the number of versions that evaluation is done on the project. F1-score is also measured for each project and the results are nearly identical to the MCC score.

The defect prediction method performance can also be measured using the \textit{precision} and \textit{recall} metric. In the context of our setting, recall is the number of bugs identified using the defect prediction model. Also, precision can be evaluated as the proportion of classes that have been correctly labeled among all classes that have been labeled as buggy by the prediction model. A bug is considered to be predicted if the corresponding class to its bug-fix has a fault-proneness higher than the project's computed threshold. This threshold is selected using the validation data to maximize the MCC score.


\begin{table}[!htb]
\captionsetup{font=normalsize}
\centering
\caption{Performance of online XGBoost on all projects}
\label{tab:dp_performance}
\begin{tabular}{M{1cm}|M{1.2cm}|M{1.5cm}|M{0.9cm}|M{1.3cm}|M{1.3cm}|M{0.9cm}}

\hline
	\textbf{Project} & \textbf{Versions} & \textbf{Evaluation versions} &  \textbf{MCC} & \textbf{F1-score} & \textbf{Precision} & \textbf{Recall} \\ \hline \hline
	Chart & 26 & 21 & 0.88 & 0.88 & 0.78 & 1  \\ \hline
	Closure & 133 & 128 & 0.89 & 0.89 & 0.90 & 0.89  \\ \hline
	Lang & 65 & 60 & 0.92 & 0.92 & 1 & 0.85  \\ \hline
	Math & 106 & 101 & 0.84 & 0.84 & 0.86 & 0.83  \\ \hline
	Time & 27 & 22 & 0.91 & 0.91 & 0.91 & 0.91  \\ \hline\hline
	\textbf{Overall} & \textbf{357} & \textbf{332} & \textbf{0.89} & \textbf{0.89} & \textbf{0.89} & \textbf{0.90}\\ \hline
\end{tabular}
\end{table}

\subsubsection{Classification model hyperparameter tuning}
To tune the hyperparameters a randomized search is done according to a distribution for the subjected parameters~\cite{zainab2020performance, sandha2020mango}. In the XGBoost classifier, the key parameters are Tree count, Max tree depth, L1 and L2 regularization coefficients, and Gamma which is the minimum change needed in the loss function to partition a leaf node. The distributions used in the randomized search are mostly uniform. To increase the flexibility of the model, the hyperparameters are tuned separately for each project. 

The next step is to make a validation set for the randomized search to select the best hyperparameters. Table~\ref{tab:dataset_partitioning} summarizes the data instances used in the hyperparameter tuning step of each project as the training, validation, and test sets. The best model is selected in terms of the Matthews correlation coefficient score \cite{yao2021impact, yao2020assessing} on the validation data. In the case of a tie, the model with the highest tree count to max depth ratio is selected.

\begin{table}[!htb]
\captionsetup{font=normalsize}
\centering
\caption{Data instances used in the hyperparameter tuning step of each project}
\label{tab:dataset_partitioning}
\begin{tabular}{c|c|c}

\hline
	\textbf{Project} & \textbf{Versions} & \textbf{Dataset}  \\ \hline \hline
	Current Project & First 5 versions & Validation \\ \hline
	Current Project & Versions higher than 5  & Test (Untouched in this step)  \\ \hline
    Other Projects & Last 5 versions  & Validation  \\ \hline
    Other Projects & First version up to the last 5 versions  & Training  \\ \hline
\end{tabular}
\end{table}

The usage of the three shrinkage strategies to minimize the risk of the curse of dimensionality discussed in Section~\ref{sec:data_preprocessing} can be verified by the parameters chosen in the hyperparameter tuning. The key parameters are Tree count, Tree max depth, Column sampling rate, Alpha, Lambda, and Gamma. Tree count is the number of trees used in the ensemble. Tree max depth is the limit within which each tree in the ensemble can grow. Hence, it is the number of features used in each tree to make the final decision. The column sampling rate is the rate of subset columns used to build each tree in the ensemble. Alpha, Lambda, and Gamma are L1/L2 regularization parameters and the minimum loss reduction of a leaf node respectively. Table~\ref{tab:dp_parameters} shows the values of the aforementioned parameters. The maximum depth for each tree alone has drastically limited the number of features that influence the decision, but in some cases, other parameters also have non-zero values which further limits the curse of dimensionality problem.

\begin{table}[!htb]
\captionsetup{font=normalsize}
\centering
\caption{Selected parameters for each project}
\label{tab:dp_parameters}
\begin{tabular}{M{1.2cm}|M{1.2cm}|M{1.2cm}|M{1.5cm}|M{0.9cm}|M{1.15cm}|M{1.15cm}}

\hline
	\textbf{Project} & \textbf{Tree count} &  \textbf{Tree max depth} & \textbf{Column sampling rate} & \textbf{Alpha} & \textbf{Lambda} & \textbf{Gamma} \\ \hline \hline
	Lang & 400 & 2 & 0.83 & 0 & 0 & 0  \\ \hline
	Math & 400 & 5 & 0.86 & 0 & 5 & 1  \\ \hline
	Chart & 300 & 4 & 0.93 & 2 & 0 & 1  \\ \hline
	Closure & 300 & 4 & 1 & 0.1 & 5 & 1  \\ \hline
	Time & 400 & 2 & 0.83 & 0 & 0 & 0  \\ \hline\hline
	\textbf{Average} & \textbf{360} & \textbf{3.4} & \textbf{0.89} & \textbf{0.42} & \textbf{2} & \textbf{0.6}\\ \hline
\end{tabular}
\end{table}

\subsubsection{Other considerations}
To overcome the imbalanced nature of the dataset, several ideas were tested. Most of the negative samples are unchanged units throughout the versions. These are more likely to be duplicate samples in terms of the final feature instances. The first idea is to remove these duplicates to reduce the negative to positive samples ratio. To further decrease this ratio, a random sub-sampling of the negative instances also takes place. These two combined have improved the final results on average. The idea of generative positive samples using SMOTE was also tested but did not further improve the average result. Therefore, only negative sampling methods were applied to balance the dataset.

\section{Results}
\label{sec:results}
In this section, we present and analyze the results of our empirical study. In this regard, we answer the research questions raised in Section~\ref{research_questions} by providing and discussing the corresponding experimental results.
\subsection{\small RQ1: Comparing traditional TCP strategies with the CovClustering method (the proposed TCP method without incorporating fault-proneness)}

\begin{figure}[!htb]
\centering
\subfloat[Chart]{\includegraphics[width=1.7in]{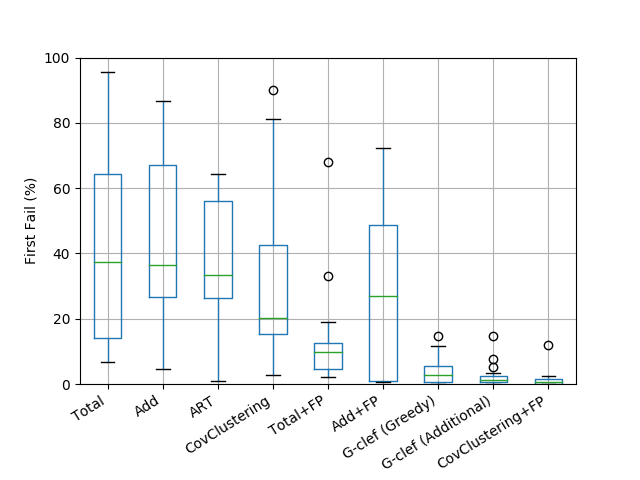}} 
\subfloat[Closure]{\includegraphics[width=1.7in]{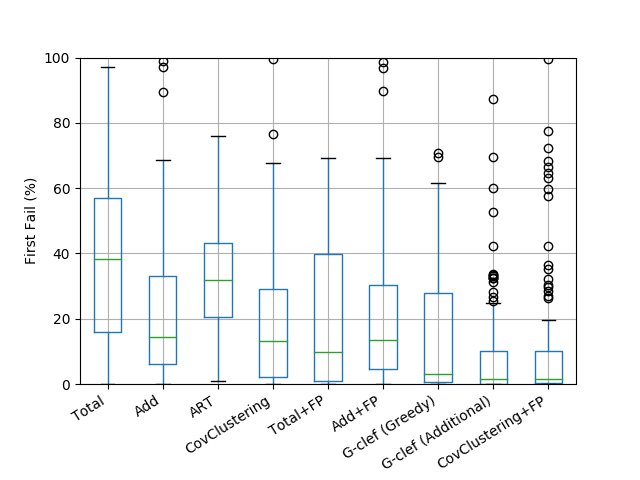}} 
\subfloat[Lang]{\includegraphics[width=1.7in]{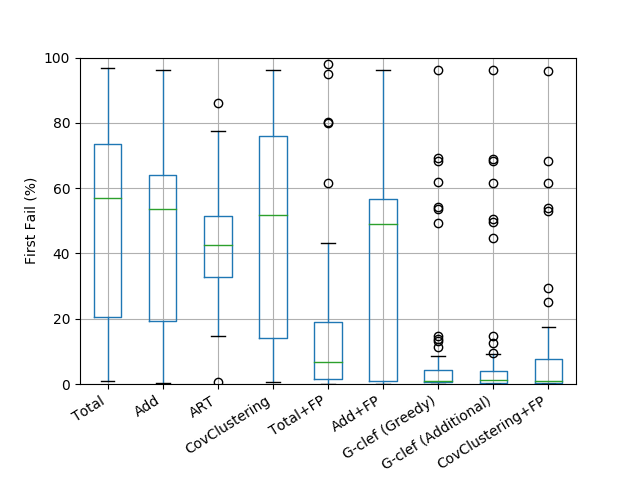}} \\
\subfloat[Math]{\includegraphics[width=1.7in]{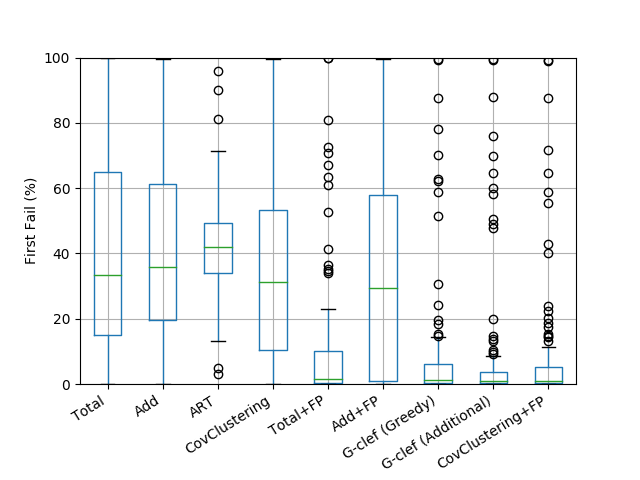}} 
\subfloat[Time]{\includegraphics[width=1.7in]{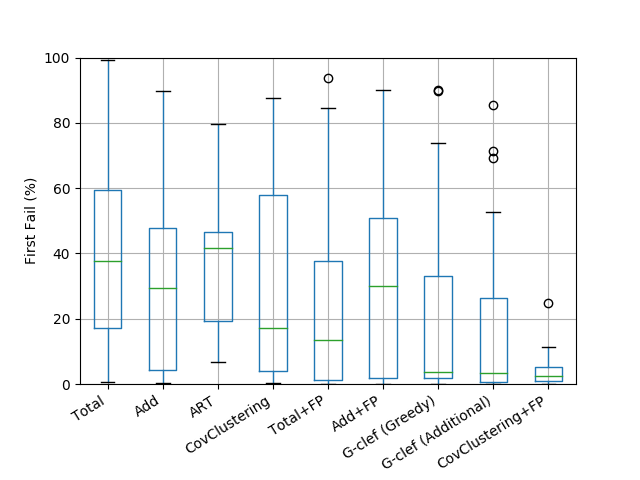}}

\caption{Evaluation results of all TCP strategies in the subject study} \label{fig:boxplots}
\end{figure}

In this research question, we compare the first-fail performance of coverage-based TCP methods with the proposed \texttt{CovClustering} TCP method (the proposed clustering-based TCP algorithm without incorporating fault-proneness). The compared methods have been described in Section~\ref{sec:subject_tcp_algorithms}. For the purpose of comparison, we have implemented and executed these methods on the subjects of study, presented in Section~\ref{sec:subjects_of_study}. The average first-fail value of these methods is shown in Table~\ref{tab:methods_firstfail}. Also, Figure~\ref{fig:boxplots} depicts the boxplot of the first-fail of these methods on the subject versions of each project. Note that lower values of first-fail mean that the TCP algorithm has detected the fault sooner, therefore algorithms with lower values of first-fail are performing better.

The \texttt{Additional} algorithm has better performance than the \texttt{Total} algorithm, which confirms previous reports~\cite{hao2015optimal}. The \texttt{ART} algorithm has interesting performance and performs better than the \texttt{Additional} algorithm on the Chart and Lang projects. However, overall their performance is near and the \texttt{Additional} algorithm  slightly performs better than the \texttt{ART} algorithm.

Our research question is related to the rows \texttt{Total}, \texttt{Additional}, \texttt{ART}, and \texttt{CovClustering} in Table~\ref{tab:methods_firstfail}. The first-fail metric of all algorithms on the Lang project are more than $40\%$ which shows that the algorithms do not perform much better than random prioritization (which should have around $50\%$ first-fail). This shows that TCP algorithms based solely on coverage, probably will not have appropriate performance on the Lang project.

As can be seen, the average first-fail metric of the \texttt{CovClustering} method is less than the other methods on all projects except the Lang project, where the \texttt{ART} method has the best performance. This comparison is also observed in the boxplots of Figure~\ref{fig:boxplots}.

The proposed \texttt{CovClustering} method has better performance on four of the five projects and is also superior on the overall value on all projects. We performed Wilcoxon signed-rank tests \cite{wilcoxon1992individual} ($p$-value $< 0.05$) to make sure that the overall superiority is statistically significant. The null hypothesis is that there is no significant difference in the first-fail performance of the \texttt{CovClustering} TCP method with respect to each of the coverage-based TCP methods. The results of this test demonstrate that there is a statistically significant difference between the  \texttt{CovClustering} TCP method with respect of other coverage-based methods: \texttt{Total} ($p$-value $ = 4.91 \times 10^{-7}$), \texttt{Additional} ($p$-value $ = 0.008$), and the \texttt{ART} ($p$-value $ = 3.52 \times 10^{-4}$). 

Therefore our answer to RQ1 is that the proposed clustering method is superior to the coverage-based TCP methods.

\makesavenoteenv{tabular}
\makesavenoteenv{table}

\begin{table}[!htb]
\captionsetup{font=normalsize}
\centering
\caption{Average first-fail scores of each TCP strategy}
\label{tab:methods_firstfail}

\begin{tabular}{c|c|c|c|c|c|c}
\hline
	\textbf{Algorithm} & Chart & Closure & Lang & Math & Time & \textbf{Overall} \\ \hline \hline
	\texttt{Total} & 41.14 & 36.04 & 50.28 & 40.32 & 41.72 & 41.90 \\ \hline
	\texttt{Additional} &  42.69 & 21.37 & 45.94 & 40.80 & 33.04 & 36.77 \\ \hline
	\texttt{ART} & 38.75 & 31.92 & \underline{41.93} & 42.37 & 36.16 & 38.23 \\ \hline
	\texttt{CovClustering} & \underline{32.07} & \underline{19.12} & 47.44 & \underline{34.87} & \underline{31.04} & \underline{32.91} \\ \hline \hline
	\texttt{Total+FP} & 13.01 & 20.20 & 15.11 & 11.32 & 25.97 & 17.12 \\ \hline
	\texttt{Additional+FP} &  30.79 & 19.52 & 37.12 & 34.24 & 30.22 & 30.38 \\ \hline
	\texttt{G-clef Original (Greedy)} & 47.9 & 47.6 & 33.1 & 31.8 & 24.7 & 37.0 \\ \hline
	\texttt{G-clef Original (Additional)} & 41.2 & 27.1 & 49.9 & 36.6 & 24.3 & 35.8 \\ \hline
	\texttt{G-clef (Greedy)} & 3.81 & 14.15 & 9.63 & 10.00 & 22.47 & 12.01 \\ \hline
	\texttt{G-clef (Additional)} & 2.40 & \textbf{\underline{9.19}} & 9.24 & 9.49 & 18.97 & 9.86 \\ \hline
	\texttt{CovClustering+FP} & \textbf{\underline{1.31}} & 10.50 & \textbf{\underline{8.79}} & \textbf{\underline{8.98}} & \textbf{\underline{4.31}} & \textbf{\underline{6.78}} \\ \hline
\end{tabular}
\end{table}

\subsection{\small RQ2: Studying the effect of incorporating fault-proneness on the proposed method}
We want to know the effect of fault-proneness on the proposed method in this research question. Therefore we compare the \texttt{CovClustering} TCP method, with the proposed \texttt{CovClustering+FP} method, in terms of the first-fail metric. The average value of the first-fail metric on the proposed method incorporating fault-proneness is shown in row \texttt{CovClustering+FP} of Table~\ref{tab:methods_firstfail}.

In addition we also compare the \texttt{CovClustering+FP} method with other state-of-the-art fault-proneness based methods which are the TCP using fault-based coverage \cite{mahdieh2020incorporating} and the G-clef algorithm \cite{paterson2019empirical}. Rows \texttt{Total+FP} and \texttt{Additional+FP} of Table~\ref{tab:methods_firstfail} correspond to the fault-based coverage TCP methods and rows \texttt{G-clef (Greedy)} and \texttt{G-clef (Additional)} correspond to the G-clef algorithm presented by Patterson et al.~\cite{paterson2019empirical}. 
To implement the fault-proneness-based algorithms we used the fault-proneness resulting from the defect prediction method proposed in this paper, to have comparable results.

\begin{table}[!htb]
\captionsetup{font=normalsize}
\centering
\caption{The details of the statistical tests of comparison of all TCP algorithms to the \texttt{CovClustering+FP} method}
\label{tab:methods_significance}
\begin{tabular}{c|c|c|c|c|c|c}
\hline
	\textbf{Algorithm} & Chart & Closure & Lang & Math & Time & \textbf{Overall} \\ \hline \hline
	\texttt{Total+FP} & $0.000^{*}$ & $0.000^{*}$ & $0.000^{*}$ & $0.044$ & $0.001$ & $0.000^{*}$ \\ \hline
	\texttt{Additional+FP} &  $0.001$ & $0.000^{*}$ & $0.000^{*}$ & $0.000^{*}$ & $0.001$ & $0.000^{*}$ \\ \hline
	\texttt{G-clef (Greedy)} & $0.001$ & $0.013$ & $0.405$ & $0.012$ & $0.068$ & $0.028$ \\ \hline
	\texttt{G-clef (Additional)} & $0.007$ & $0.540$ & $0.587$ & $0.044$ & $0.040$ & $0.062$ \\ \hline
\end{tabular}
    \begin{tablenotes}
    \footnotesize
    \item ($0.000^{*}$ denotes values less than 0.001 which are typically very small)
    \end{tablenotes}
\end{table}

We have also provided the original results of the G-clef algorithm noted in their paper~\cite{paterson2019empirical} as two rows of Table~\ref{tab:methods_firstfail}, as they have used the same initial dataset used by our study (the Defects4J dataset). It is observable in Table~\ref{tab:methods_firstfail} that the performance of the G-clef algorithm using the fault-proneness results of this paper is significantly better than the G-clef results in the original paper. This observation confirms that the defect prediction method proposed in this paper has remarkable performance.

It is observed that the algorithms based on fault-proneness (the lower part of Table~\ref{tab:methods_firstfail}) have better performance than the purely coverage-based algorithms (the upper part of Table~\ref{tab:methods_firstfail}). The performance of \texttt{Total+FP} is also interesting and it is superior to \texttt{Additional+FP}, unlike their purely coverage-based counterparts. This shows that the highly competitive \texttt{Additional} algorithm will not improve enough when naively applying fault-proneness to the coverage formulation.

Comparing the results presented in Table~\ref{tab:methods_firstfail}, it is observed that the \texttt{CovClustering+FP} method has the best value of average first-fail compared to other TCP algorithms, in all but one case. The only exception to this observation is the comparison of the \texttt{G-clef (Additional)} algorithm on the Closure project.

We performed the Wilcoxon signed-rank again to evaluate the significance of the results. The null hypothesis is that there is no significant difference in the first-fail performance of the \texttt{CovClustering+FP} method with respect to the other TCP algorithms. The results of this test are shown in Table~\ref{tab:methods_significance}. The null hypothesis is rejected in cases where the values are less than 0.05, and in these cases, there is a significant difference between the proposed \texttt{CovClustering+FP} method and other algorithms. The very low $p$-values indicate that significance is confident. For most of the table, this significance is verified however the statistical test fails for some of the values relating to the G-clef methods, specifically on the Closure and Lang project. 

The conclusion is that the \texttt{CovClustering+FP} performs better than the \texttt{Total+FP} and \texttt{Additional+FP} algorithms of \cite{mahdieh2020incorporating} but does not significantly dominate the \texttt{G-clef (Additional)} algorithm in all cases. Note that as mentioned and observed in Table~\ref{tab:methods_firstfail}, \texttt{CovClustering+FP} performs much better than the original \texttt{G-clef} implementation and the presented results are due to boosting \texttt{G-clef} with the defect prediction method presented here.

\subsection{\small RQ3: The effect of the clustering configurations (distance function and number of clusters) on the effectiveness of the proposed TCP strategies}
\label{sec:rq3_answer}
In this research question, we study the effect of the distance function and number of clusters on the effectiveness of the \texttt{CovClustering} and \texttt{CovClustering+FP} methods. In this manner, we experiment the proposed methods, with different distance functions and vary the number of clusters in a specified range.

\subsubsection{Distance function}
We experiment using three well-known distance functions which have been also used in previous related research~\cite{pan2022test}: Euclidean distance, Manhattan distance, and distance based on Cosine similarity.

\begin{figure}[!htb]
\centering
\subfloat{\includegraphics[width=1.7in]{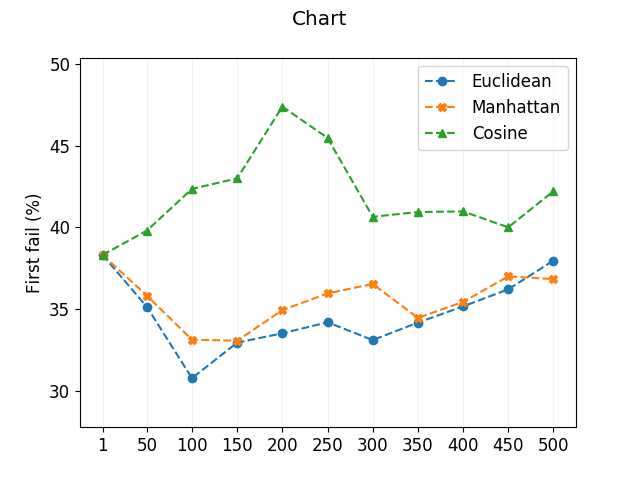}} 
\subfloat{\includegraphics[width=1.7in]{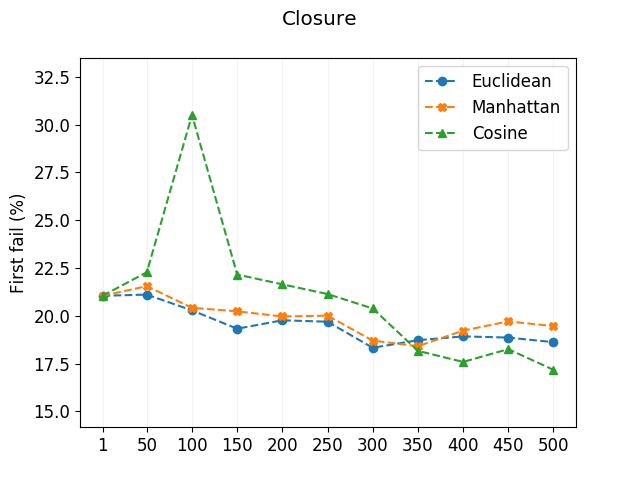}} 
\subfloat{\includegraphics[width=1.7in]{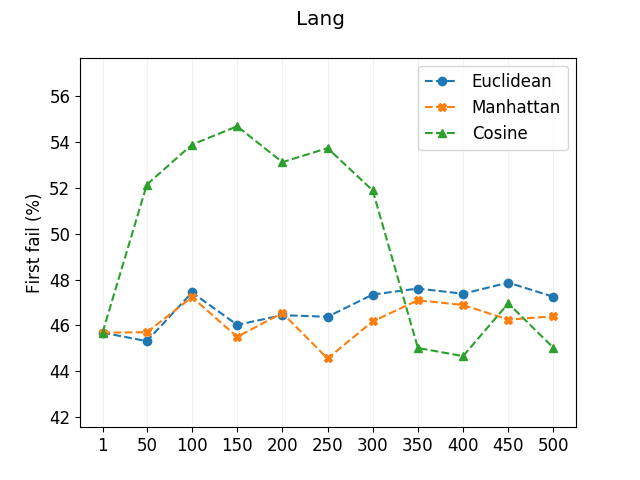}} \\
\subfloat{\includegraphics[width=1.7in]{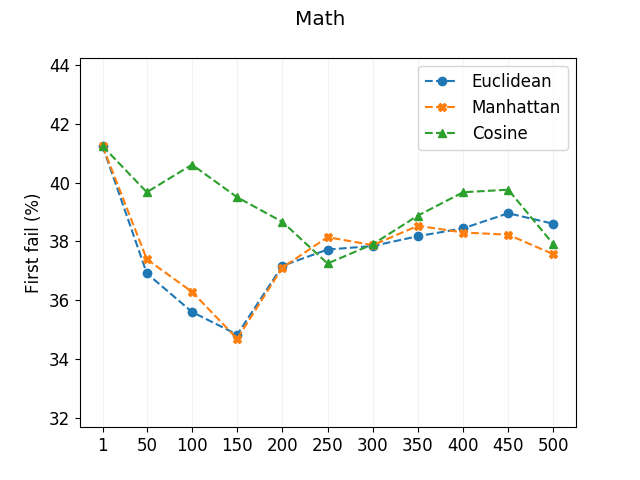}} 
\subfloat{\includegraphics[width=1.7in]{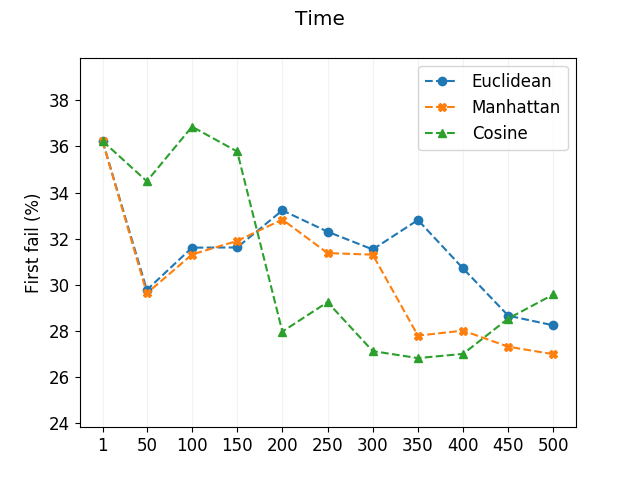}}

\caption{The performance of the proposed \texttt{CovClustering} TCP strategy on the subject study using different distance functions and cluster numbers (RQ3, Distance function)} \label{fig:clustering_np_fp}
\end{figure}

\begin{figure}[!htb]
\centering
\subfloat{\includegraphics[width=1.7in]{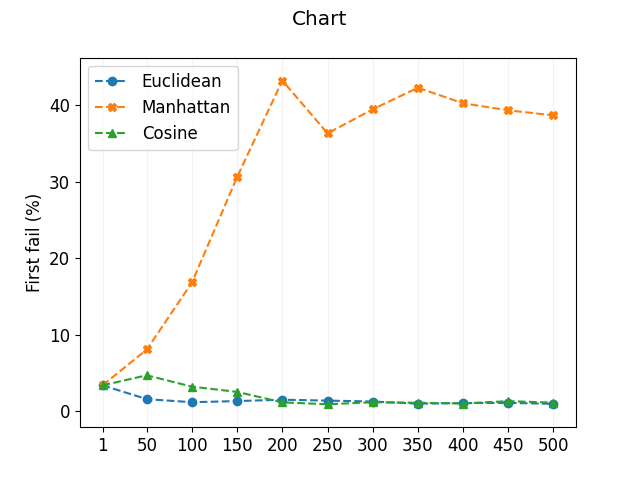}} 
\subfloat{\includegraphics[width=1.7in]{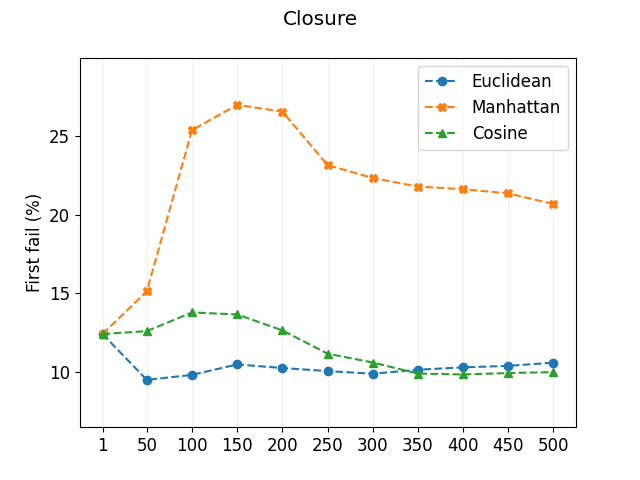}} 
\subfloat{\includegraphics[width=1.7in]{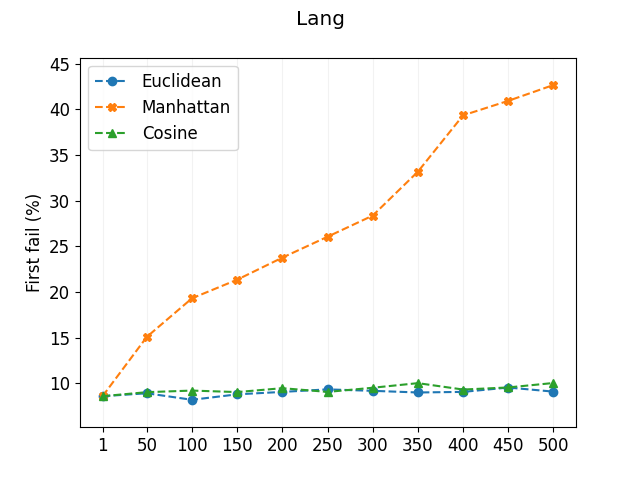}} \\
\subfloat{\includegraphics[width=1.7in]{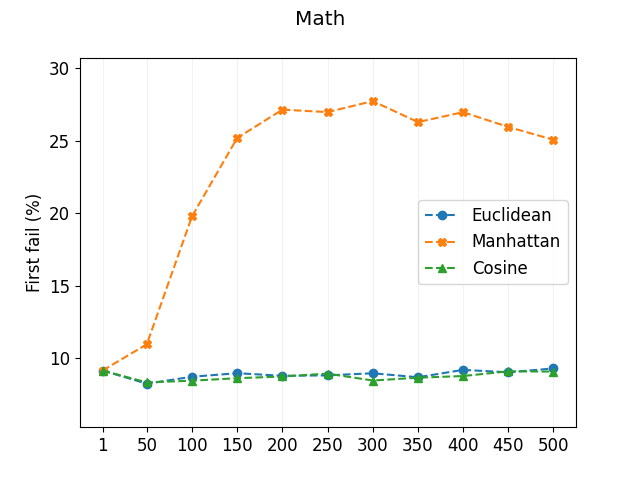}} 
\subfloat{\includegraphics[width=1.7in]{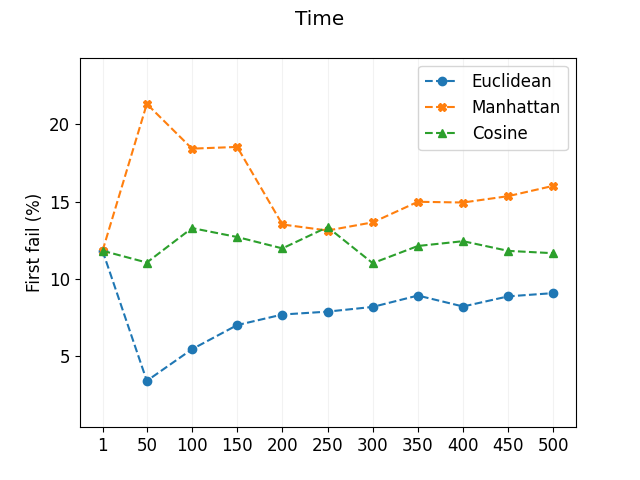}}

\caption{The performance of the proposed \texttt{CovClustering+FP} TCP strategy on the subject study using different distance functions and cluster numbers (RQ3, Distance function)} \label{fig:clustering_max_fp}
\end{figure}

Figure~\ref{fig:clustering_np_fp} and \ref{fig:clustering_max_fp} show the performance of executing the \texttt{CovClustering} and \texttt{CovClustering+FP} method through different distance functions and cluster numbers. For \texttt{CovClustering} the Euclidean and Manhattan distance have similar performance and seem to have better performance than cosine similarity on most points. However, for the \texttt{CovClustering+FP} method, Euclidean distance shows competitive performance compared to other distance functions on all subject projects. The superiority of Euclidean distance for diversity-based TCP algorithms has also been observed by other researchers~\cite{wang2016empirical}. Therefore we can hope that generally, Euclidean distance can be more appropriate for this application as a first choice, but experimenting with other distance functions can also be considered. This also shows that the accuracy of defect prediction can highly impact the performance of fault-proneness methods.

\subsubsection{Number of clusters}
\label{sec:rq_num_of_clusters}
To choose the appropriate number of clusters, a well-practiced technique is to employ metrics that evaluate the quality clustering to get a better insight. Among these metrics, we utilize the \textit{Davies–Bouldin index (DBI)} \cite{davies1979cluster}. DBI is defined as the average ratio of within-cluster distances of each cluster to the between-cluster distances to the nearest cluster. Thus, more compact clusters will result in a better score. Lower values of DBI indicate better clustering, and therefore the range of points that have low values of DBI are candidates for choosing the number of clusters. However, we must consider some tips in choosing the number of clusters e.g. we shouldn't consider the points in the steady slope at the end of the curve. Using this approach we show an appropriate number of clusters intuitively chosen considering the DBI metric in Table~\ref{tab:optimal_cluster}.

\begin{table}[!htb]
\captionsetup{font=normalsize}
\centering
\caption{Best numbers of clusters chosen heuristically using the DBI metric}
\label{tab:optimal_cluster}
\begin{tabular}{P{1.5cm}|P{3.5cm}|P{4cm}}
\hline
	\textbf{Project} & \texttt{CovClustering} best cluster number by DBI & \texttt{Covclustering+FP} best cluster number by DBI \\ \hline \hline
	Chart & 100 & 175 \\ \hline
	Closure & 150 & 150 \\ \hline
	Lang & 175 & 150 \\ \hline
	Math & 125 & 150 \\ \hline
	Time & 150 & 75 \\ \hline \hline
\end{tabular}
\end{table}

\begin{figure}[!htb]
\centering
\subfloat{\includegraphics[width=1.7in]{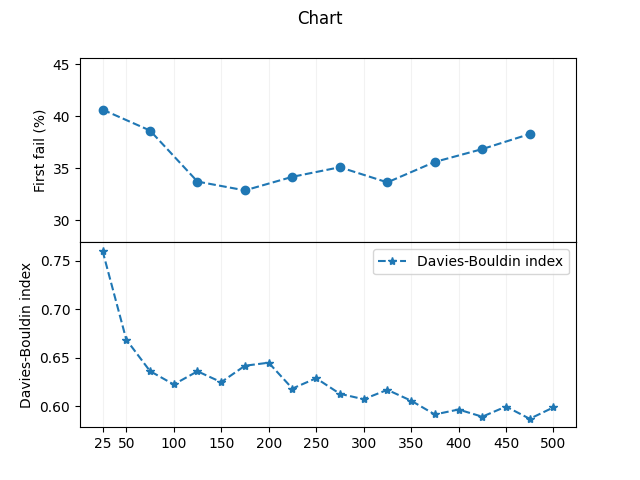}} 
\subfloat{\includegraphics[width=1.7in]{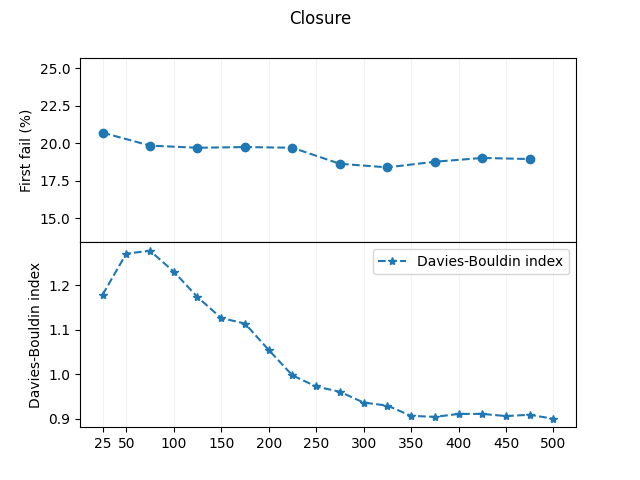}} 
\subfloat{\includegraphics[width=1.7in]{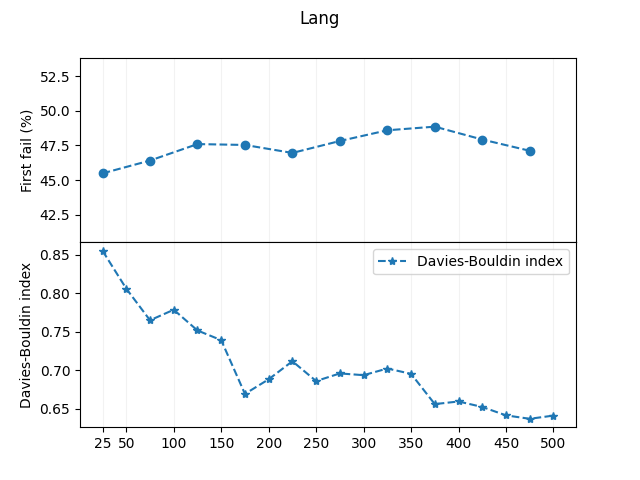}} \\
\subfloat{\includegraphics[width=1.7in]{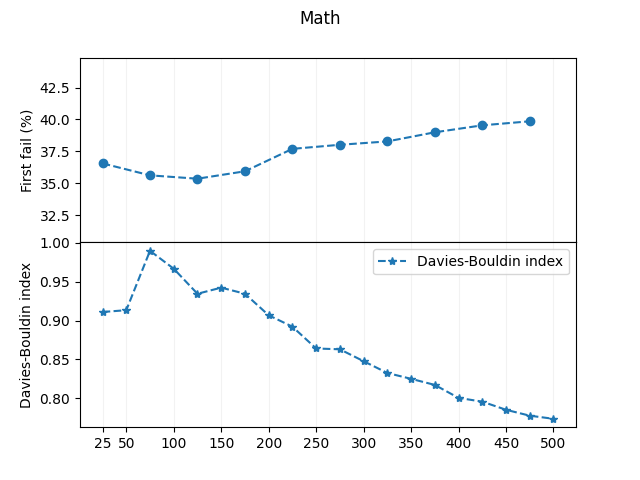}} 
\subfloat{\includegraphics[width=1.7in]{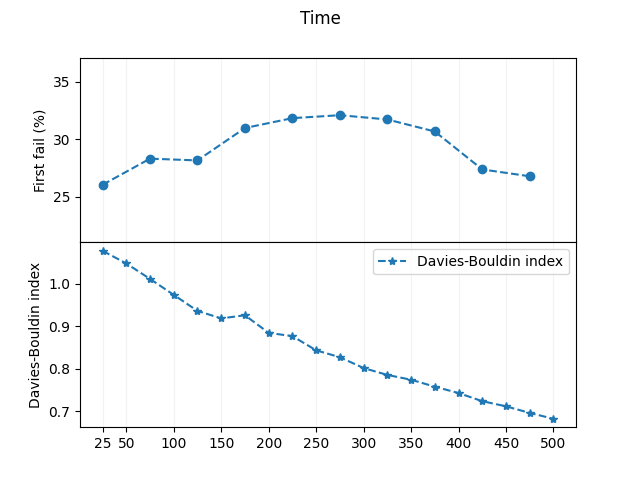}}

\caption{The DBI value of the clustering of the \texttt{CovClustering} method on the subject projects using different cluster numbers (RQ3, Number of clusters)} \label{fig:metrics_nofp}
\end{figure}

\begin{figure}[!htb]
\centering
\subfloat{\includegraphics[width=1.7in]{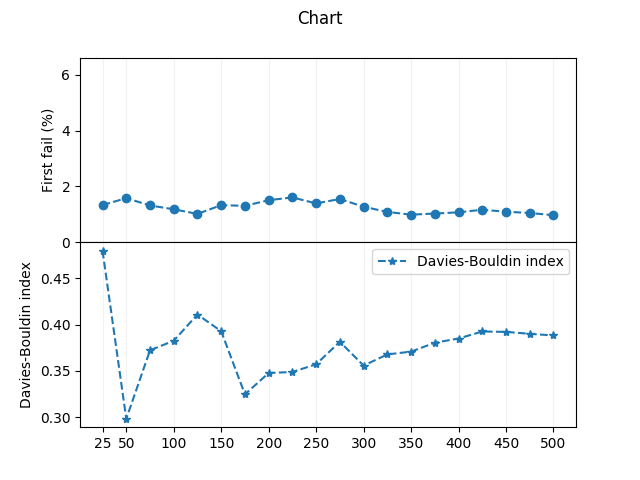}} 
\subfloat{\includegraphics[width=1.7in]{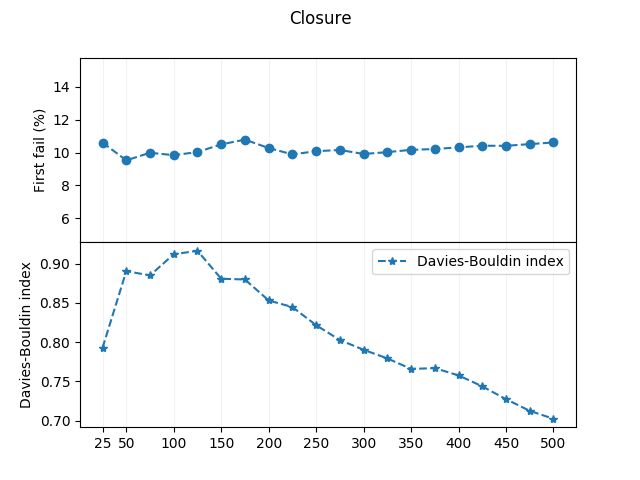}} 
\subfloat{\includegraphics[width=1.7in]{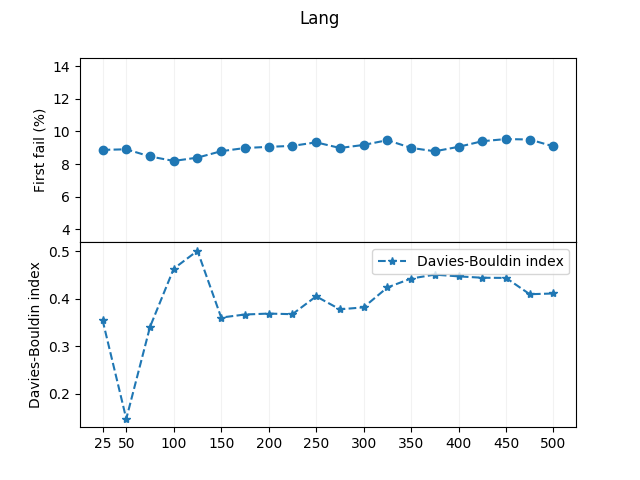}} \\
\subfloat{\includegraphics[width=1.7in]{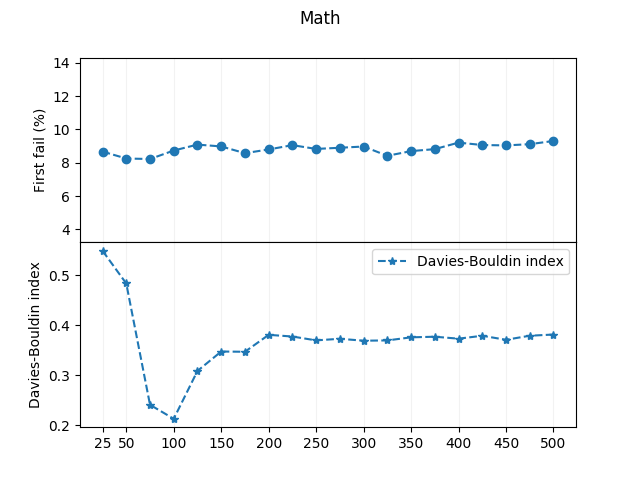}} 
\subfloat{\includegraphics[width=1.7in]{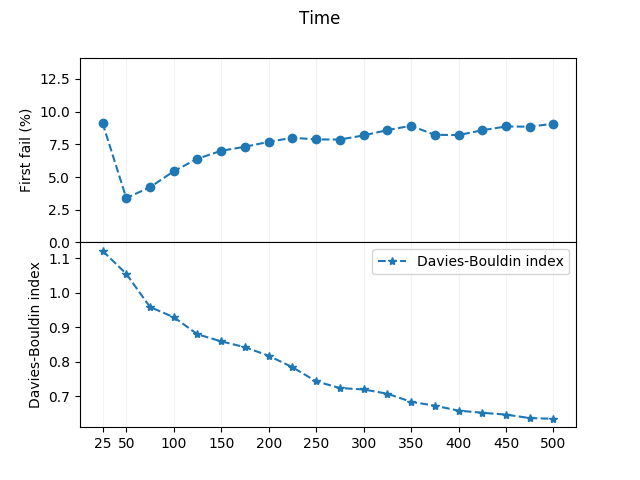}}

\caption{The DBI value of the clustering of the \texttt{CovClustering+FP} method on the subject projects using different cluster numbers (RQ3, Number of clusters)}
\label{fig:metrics_fp}
\end{figure}

Additionally, to validate this approach and also observe the effect of the number of clusters on the proposed strategies, we show the result of experimentation of the proposed methods with a varied number of clusters in the range of $[25, 500]$, in Figure~\ref{fig:metrics_nofp} and Figure~\ref{fig:metrics_fp}. It is observed that the chosen number of clusters has relatively good TCP performance, confirming the chosen approach. Furthermore, it is observed in Figure~\ref{fig:metrics_fp} that the number of clusters has less impact on the performance of the \texttt{CovClustering+FP} method on the subject projects when compared to \texttt{CovClustering}.

\section{Discussion}
\label{sec:discussion}
\subsection{Practical Considerations}
To apply the proposed approach, the coverage of the test suite must be measured. This measurement can be done once and used for subsequent changes to the source code until a certain point. This shortcut technique is also applicable for the defect prediction phase and also distance computation. The coverage measurement can be implemented using static analysis methods to speed up this process. The granularity of code units we have experimented with for coverage measurement is method-level, but measuring coverage in statement-level can be even more effective.

There were no theoretical assumptions about the defect prediction phase, therefore it can be replaced with any standard defect prediction method with hopefully appropriate results. Additionally, The proposed classification process can be used with other feature sets extracted from the source code. Applying  cross-project methods can help the usage of the proposed method in the early stage of a project.

One important practical advantage of the proposed TCP method is that the proposed method is designed such that using a weak classifier for the defect prediction phase does not deteriorate the whole TCP result. This is due to the clustering phase which is independent of the fault-proneness estimations and provides a stable base platform for test-case diversification.

\subsection{Threats to validity}
\textit{Construct Validity.}
Construct validity focuses on the relation between the theory behind the experiment and the observed results. One of the main concerns of this threat is related to the evaluation metrics in our experiment. We considered the first-fail metrics as measures of the effectiveness of test case prioritization. The first failing metric has also been previously used \cite{noor2015similarity, noor2017studying, palma2018improvement, rahman2018prioritizing, abou2021detrimental} and is also reasonable to be used when only a few test cases fail such as our case.

\textit{Internal Validity.}
Internal validity refers to whether the relationship between the experiment itself and the result obtained is causal rather than the result of other factors. A major part of our experiment concerns the Defects4J and Defects4J+M datasets which we have relied on because of being previously reviewed by other researchers~\cite{paterson2019empirical,noor2015similarity,luo2018assessing,mahdieh2020incorporating}. The main methods of our implementation are also parts of standard libraries which have been thoroughly tested. One concern that can be mentioned is using some parameters in methods, such as the number of clusters. In practice, choosing these parameters can be done using heuristics (which is practiced in this paper) or by checking multiple values and choosing the one with the best performance. 

\textit{External Validity.}
The experimental procedure has been performed on projects which are mostly implemented in the Java language; therefore, the results might differ in projects developed using other languages. However, the clustering algorithms are completely language-neutral and the defect prediction procedure is mostly based on language-independent features.
Additionally, by using popular open-source projects that contain large test suites in our empirical study, we tried to study a completely real-world scenario. Despite these facts, in the future, we have to evaluate our approaches using projects with different languages and other characteristics to ensure that the results are generalizable.

\textit{Conclusion Validity.}
Conclusion validity focuses on the significance of the treatment specifically the statistical validity of the conclusions. To enhance conclusion validity, we applied the Wilcoxon signed-rank statistical tests to the results of the experiments to validate the significance of the conclusions of comparing the performance of the algorithms with each other. The Wilcoxon signed-rank has been applied in many TCP experimental comparisons \cite{luo2016large, kwon2014test, wang2017qtep, luo2018static}. This non-parametric test is chosen since we did not make assumptions that the data under consideration is normally distributed.

\section{Related Work}
\label{sec:related_work}
Much research has been conducted in the last two decades to study different methods and analyze their performance for test case prioritization in the context of regression testing. From a big-picture point of view, TCP methods can be categorized based on two aspects: different sources of information used for TCP and various heuristics and optimization strategies used for ordering the test cases~\cite{hemmati2019advances}. We continue by reviewing different TCP algorithms, considering their source of information and optimization strategy.

Considering the first point of view, many TCP studies have used code coverage as a major source of information for prioritization~\cite{khatibsyarbini2018test}. These methods are based on the assumption that test cases with larger coverage have a better ability for fault detection. Other sources of information have also been used for TCP such as historical failure data of test cases, formal specifications or requirements, and source code metrics~\cite{hemmati2019advances}.

Lachman et al.~\cite{lachmann2016system} propose using  machine learning techniques to leverage test case execution history and test case description texts for prioritizing manual system-level test cases. Their method runs in a completely black-box context which implies better applicability of the method in practice.
Hettiarachchi et al.~\cite{hettiarachchi2016risk} designed a fuzzy expert system which  estimates the risks of system requirements and then prioritize test cases based on the risks which they cover. Arafeen and Do \cite{arafeen2013test} propose a method that first clusters the requirements based on a text-mining technique and then uses the requirements-test cases traceability matrix to cluster the test cases. Afterward, the test cases in each cluster are prioritized using source code metrics such as McCabe Cyclomatic Complexity and finally, the test cases are prioritized according to the importance of the requirements to the clients.

Noor et al.~\cite{noor2015similarity} propose a similarity-based TCP approach based on historical failure data of test cases. Their method uses the intuition that test cases that are similar to failed test cases in the past are probable of fault detection. 


As another source of information, several researchers have been proposed methods to utilize bug history for test case selection and prioritization. Laali et al. \cite{laali2016test} propose an online TCP method that utilizes the locations of faults in the source code revealed by failed test cases to prioritize the non-executed test cases. A method utilizing previously fixed faults to choose a small set of test cases for test selection is proposed by Engstrom et al. \cite{engstrom2010empirical}. Some studies such as \cite{anderson2014improving} and  \cite{engstrom2011improving}, suggest the idea of using the failure history of regression test cases to improve future regression testing phases. Kim et al. \cite{kim2010effective} borrowed ideas from fault localization to tackle the TCP problem. By considering the fact that defects are fixed after being detected, they guess that test cases covering previous faults must have lower priority in TCP ordering because they will have lower fault detection possibility.
Wang et al. \cite{wang2017qtep} proposed a quality-aware TCP method (QTEP) that uses static bug finders and unsupervised methods for defect prediction. Paterson et al. \cite{paterson2019empirical} proposed a ranked-based technique to prioritize test cases that estimate the likelihood of java classes having bugs. Their experiments show that using their TCP method reduces the number of test cases required to find a fault compared with existing coverage-based strategies.

Multi-objective evolutionary techniques have been of interest for TCP and test selection as they can tackle two or more different kinds of objectives (such as code coverage, requirements coverage, etc.) for prioritization~\cite{wang2016enhancing, yoo2007pareto, mondal2015exploring}. Pradhan et al.~\cite{pradhan2019employing} employ rule mining on the test execution history to extract relationships among test cases and use multi-objective algorithms to prioritize test cases in a black-box setting.

Due to the relatively high computation cost of TCP algorithms, proposing TCP methods with lower computation costs for large-scale test suites has been investigated. Miranda et al.~\cite{miranda2018fast} propose using hashing-based approaches to provide faster TCP algorithms.

Reinforcement Learning (RL) based Continuous Integration (CI) testing \cite{spieker2017reinforcement} was introduced to prioritize test cases based on applying reinforcement learning techniques to the test case execution history of continuous integration systems. Bagherzadeh et al.~\cite{bagherzadeh2021reinforcement} provide RL-based TCP methods for CI, employing both the test execution history and light-weight code features and show that their methods are effective.

We will thoroughly review methods that utilized clustering methods for test case diversification, due to their relation to the approach of this paper. Carlson et al. \cite{carlson2011clustering} proposed a method based on clustering of method coverage for TCP. Their approach works in two steps, in the first step the test cases are clustered using code coverage similarity. The clustering is performed using an agglomerative hierarchical clustering method \cite{tan2016introduction}. In the second step, the test cases of each cluster are prioritized using multiple metrics such as code coverage, code complexity, fault history, and a combination of these metrics. They empirically investigate their method on a subset of the Microsoft Dynamics AX project. Their results show that using clustering improves TCP compared to prioritizing without clustering. By utilizing the same coverage clustering method and prioritizing test cases according to the previous failure history of test cases, another TCP method is proposed by Fu et al. \cite{fu2017coverage}. Their method also uses estimations of failure rate according to the program line changes.

Chen et al. \cite{chen2018test} employed ideas from adaptive random testing and clustering to propose TCP methods for object-oriented software. Their methods start by clustering the test cases by comparing the number of objects and methods and also the Object and Method Invocation Sequence Similarity (OMISS) metric \cite{chen2016similarity}. Afterward, the clusters are sorted in an adaptive random sequence and test cases are sampled iteratively according to the order of clusters. Zhao et al. \cite{zhao2015clustering} have combined Bayesian networks with coverage-based clustering for TCP. Their method works by prioritizing test cases in each cluster by the Bayesian network proposed by Mirarab and Tahvildari \cite{mirarab2007prioritization} which uses change information, software quality metrics, and test coverage as data sources. They conclude that their TCP method has a higher fault detection rate than the plain Bayesian network-based approach of~\cite{mirarab2007prioritization}.

Fang et al. \cite{fang2014similarity} introduced a new test case similarity measure by comparing the ordering of execution count on program entities. They use this similarity measure to prioritize test cases using both an adaptive random testing inspired method and a clustering-based method. These methods are evaluated empirically by creating mutant versions of multiple open-source projects and measuring the fault detection rate.

History of test case failure is used by Abu Hasan et al.~\cite{hasan2017test} to improve clustering-based TCP. Their proposed methods order test cases in order of similarity to test cases that have failed in previous versions. Their empirical study is based on real-world faults however the number of versions experimented with is very limited (only 2 versions of 3 projects) and the comparison of their methods is done with random TCP.

A test case failure prediction method based on coverage clustering is proposed by Pang et al. \cite{pang2013identifying}. Their approach divides test cases into two categories of effective and ineffective through $k$-means clustering. Their results show that their coverage clustering method is effective in failure prediction. This result confirms our assumption that test cases with similar execution coverage are likely to have similar failure detection capability.


\section{Conclusions and future work}
\label{sec:conclusions}
In this paper, to address the challenges of test case prioritization, we propose a method that combines the ideas of test case diversification and the incorporation of fault-proneness estimations. Specifically, we leverage defect prediction models to estimate the fault-proneness of source code areas and use agglomerative clustering to diversify the test cases. The difference between the proposed method with other state-of-the-art TCP methods is that it considers fault-proneness and diversification at the same time in a natural composition. The method proposed can also be extended to other scenarios such as other types of information sources for diversification.

We conducted an empirical study on 357 versions of five real-world projects included in the Defects4J dataset to investigate and compare different approaches. Our evaluation shows that the proposed clustering-based TCP methods are a great improvement over traditional coverage-based TCP methods. Also, the proposed combination of clustering and fault-proneness for TCP is superior to the naive fault-proneness-based TCP methods.

In future work, it is possible to apply the proposed techniques for other software testing applications such as automatic test case generation, test suite reduction, and test selection. Also, in the internal and final test case ordering phases we have used specific strategies, but other strategies can also be studied. To further study and evaluate these methods, these approaches can be executed on other subject programming languages and datasets. Furthermore, applying other methods for defect prediction, such as unsupervised methods can be interesting.

\section*{Acknowledgments}
The authors would like to thank Mrs. Monireh Ghasri for her helpful comments and tips and Dr. Hamid Beigi, Mr. Ahmad Khajehnejad, and members of the Intelligent Systems Laboratory (ISL) for kindly providing hardware facilities.

\section*{Declarations}
\textbf{Funding} Not applicable. \\\\
\textbf{Conflict of interests} The author(s) declare(s) that there is no conflict of interest regarding the publication of this manuscript.

\bigskip
\begin{appendices}




\end{appendices}


\bibliography{sn-article}


\begin{thebibliography}{146}
\ifx \bisbn   \undefined \def \bisbn  #1{ISBN #1}\fi
\ifx \binits  \undefined \def \binits#1{#1}\fi
\ifx \bauthor  \undefined \def \bauthor#1{#1}\fi
\ifx \batitle  \undefined \def \batitle#1{#1}\fi
\ifx \bjtitle  \undefined \def \bjtitle#1{#1}\fi
\ifx \bvolume  \undefined \def \bvolume#1{\textbf{#1}}\fi
\ifx \byear  \undefined \def \byear#1{#1}\fi
\ifx \bissue  \undefined \def \bissue#1{#1}\fi
\ifx \bfpage  \undefined \def \bfpage#1{#1}\fi
\ifx \blpage  \undefined \def \blpage #1{#1}\fi
\ifx \burl  \undefined \def \burl#1{\textsf{#1}}\fi
\ifx \doiurl  \undefined \def \doiurl#1{\url{https://doi.org/#1}}\fi
\ifx \betal  \undefined \def \betal{\textit{et al.}}\fi
\ifx \binstitute  \undefined \def \binstitute#1{#1}\fi
\ifx \binstitutionaled  \undefined \def \binstitutionaled#1{#1}\fi
\ifx \bctitle  \undefined \def \bctitle#1{#1}\fi
\ifx \beditor  \undefined \def \beditor#1{#1}\fi
\ifx \bpublisher  \undefined \def \bpublisher#1{#1}\fi
\ifx \bbtitle  \undefined \def \bbtitle#1{#1}\fi
\ifx \bedition  \undefined \def \bedition#1{#1}\fi
\ifx \bseriesno  \undefined \def \bseriesno#1{#1}\fi
\ifx \blocation  \undefined \def \blocation#1{#1}\fi
\ifx \bsertitle  \undefined \def \bsertitle#1{#1}\fi
\ifx \bsnm \undefined \def \bsnm#1{#1}\fi
\ifx \bsuffix \undefined \def \bsuffix#1{#1}\fi
\ifx \bparticle \undefined \def \bparticle#1{#1}\fi
\ifx \barticle \undefined \def \barticle#1{#1}\fi
\bibcommenthead
\ifx \bconfdate \undefined \def \bconfdate #1{#1}\fi
\ifx \botherref \undefined \def \botherref #1{#1}\fi
\ifx \url \undefined \def \url#1{\textsf{#1}}\fi
\ifx \bchapter \undefined \def \bchapter#1{#1}\fi
\ifx \bbook \undefined \def \bbook#1{#1}\fi
\ifx \bcomment \undefined \def \bcomment#1{#1}\fi
\ifx \oauthor \undefined \def \oauthor#1{#1}\fi
\ifx \citeauthoryear \undefined \def \citeauthoryear#1{#1}\fi
\ifx \endbibitem  \undefined \def \endbibitem {}\fi
\ifx \bconflocation  \undefined \def \bconflocation#1{#1}\fi
\ifx \arxivurl  \undefined \def \arxivurl#1{\textsf{#1}}\fi
\csname PreBibitemsHook\endcsname

\bibitem{kumar2010development}
\begin{botherref}
\oauthor{\bsnm{Kumar}, \binits{A.}}:
Development at the speed and scale of google.
QCon San Francisco
(2010)
\end{botherref}
\endbibitem

\bibitem{memon2017taming}
\begin{bchapter}
\bauthor{\bsnm{Memon}, \binits{A.}},
\bauthor{\bsnm{Gao}, \binits{Z.}},
\bauthor{\bsnm{Nguyen}, \binits{B.}},
\bauthor{\bsnm{Dhanda}, \binits{S.}},
\bauthor{\bsnm{Nickell}, \binits{E.}},
\bauthor{\bsnm{Siemborski}, \binits{R.}},
\bauthor{\bsnm{Micco}, \binits{J.}}:
\bctitle{Taming google-scale continuous testing}.
In: \bbtitle{39th International Conference on Software Engineering: Software
  Engineering in Practice Track (ICSE-SEIP)},
pp. \bfpage{233}--\blpage{242}
(\byear{2017}).
\bcomment{IEEE}
\end{bchapter}
\endbibitem

\bibitem{zhong2008experimental}
\begin{barticle}
\bauthor{\bsnm{Zhong}, \binits{H.}},
\bauthor{\bsnm{Zhang}, \binits{L.}},
\bauthor{\bsnm{Mei}, \binits{H.}}:
\batitle{An experimental study of four typical test suite reduction
  techniques}.
\bjtitle{Information and Software Technology}
\bvolume{50}(\bissue{6}),
\bfpage{534}--\blpage{546}
(\byear{2008})
\end{barticle}
\endbibitem

\bibitem{fraser2007redundancy}
\begin{bchapter}
\bauthor{\bsnm{Fraser}, \binits{G.}},
\bauthor{\bsnm{Wotawa}, \binits{F.}}:
\bctitle{Redundancy based test-suite reduction}.
In: \bbtitle{International Conference on Fundamental Approaches to Software
  Engineering},
pp. \bfpage{291}--\blpage{305}
(\byear{2007}).
\bcomment{Springer}
\end{bchapter}
\endbibitem

\bibitem{chen2017assertions}
\begin{bchapter}
\bauthor{\bsnm{Chen}, \binits{J.}},
\bauthor{\bsnm{Bai}, \binits{Y.}},
\bauthor{\bsnm{Hao}, \binits{D.}},
\bauthor{\bsnm{Zhang}, \binits{L.}},
\bauthor{\bsnm{Zhang}, \binits{L.}},
\bauthor{\bsnm{Xie}, \binits{B.}}:
\bctitle{How do assertions impact coverage-based test-suite reduction?}
In: \bbtitle{2017 IEEE International Conference on Software Testing,
  Verification and Validation (ICST)},
pp. \bfpage{418}--\blpage{423}
(\byear{2017}).
\bcomment{IEEE}
\end{bchapter}
\endbibitem

\bibitem{grindal2006evaluation}
\begin{barticle}
\bauthor{\bsnm{Grindal}, \binits{M.}},
\bauthor{\bsnm{Lindstr{\"o}m}, \binits{B.}},
\bauthor{\bsnm{Offutt}, \binits{J.}},
\bauthor{\bsnm{Andler}, \binits{S.F.}}:
\batitle{An evaluation of combination strategies for test case selection}.
\bjtitle{Empirical Software Engineering}
\bvolume{11}(\bissue{4}),
\bfpage{583}--\blpage{611}
(\byear{2006})
\end{barticle}
\endbibitem

\bibitem{yoo2007pareto}
\begin{bchapter}
\bauthor{\bsnm{Yoo}, \binits{S.}},
\bauthor{\bsnm{Harman}, \binits{M.}}:
\bctitle{Pareto efficient multi-objective test case selection}.
In: \bbtitle{Proceedings of the 2007 International Symposium on Software
  Testing and Analysis},
pp. \bfpage{140}--\blpage{150}
(\byear{2007})
\end{bchapter}
\endbibitem

\bibitem{kazmi2017effective}
\begin{barticle}
\bauthor{\bsnm{Kazmi}, \binits{R.}},
\bauthor{\bsnm{Jawawi}, \binits{D.N.}},
\bauthor{\bsnm{Mohamad}, \binits{R.}},
\bauthor{\bsnm{Ghani}, \binits{I.}}:
\batitle{Effective regression test case selection: A systematic literature
  review}.
\bjtitle{ACM Computing Surveys (CSUR)}
\bvolume{50}(\bissue{2}),
\bfpage{1}--\blpage{32}
(\byear{2017})
\end{barticle}
\endbibitem

\bibitem{jones2003test}
\begin{barticle}
\bauthor{\bsnm{Jones}, \binits{J.A.}},
\bauthor{\bsnm{Harrold}, \binits{M.J.}}:
\batitle{Test-suite reduction and prioritization for modified
  condition/decision coverage}.
\bjtitle{IEEE Transactions on software Engineering}
\bvolume{29}(\bissue{3}),
\bfpage{195}--\blpage{209}
(\byear{2003})
\end{barticle}
\endbibitem

\bibitem{elbaum2002test}
\begin{barticle}
\bauthor{\bsnm{Elbaum}, \binits{S.}},
\bauthor{\bsnm{Malishevsky}, \binits{A.G.}},
\bauthor{\bsnm{Rothermel}, \binits{G.}}:
\batitle{Test case prioritization: A family of empirical studies}.
\bjtitle{IEEE transactions on software engineering}
\bvolume{28}(\bissue{2}),
\bfpage{159}--\blpage{182}
(\byear{2002})
\end{barticle}
\endbibitem

\bibitem{lou2019survey}
\begin{bchapter}
\bauthor{\bsnm{Lou}, \binits{Y.}},
\bauthor{\bsnm{Chen}, \binits{J.}},
\bauthor{\bsnm{Zhang}, \binits{L.}},
\bauthor{\bsnm{Hao}, \binits{D.}}:
\bctitle{A survey on regression test-case prioritization}.
In: \bbtitle{Advances in Computers}
vol. \bseriesno{113},
pp. \bfpage{1}--\blpage{46}.
\bpublisher{Elsevier},
\blocation{Amsterdam, The Netherlands}
(\byear{2019})
\end{bchapter}
\endbibitem

\bibitem{khatibsyarbini2018test}
\begin{barticle}
\bauthor{\bsnm{Khatibsyarbini}, \binits{M.}},
\bauthor{\bsnm{Isa}, \binits{M.A.}},
\bauthor{\bsnm{Jawawi}, \binits{D.N.}},
\bauthor{\bsnm{Tumeng}, \binits{R.}}:
\batitle{Test case prioritization approaches in regression testing: A
  systematic literature review}.
\bjtitle{Information and Software Technology}
\bvolume{93},
\bfpage{74}--\blpage{93}
(\byear{2018})
\end{barticle}
\endbibitem

\bibitem{hemmati2019advances}
\begin{bchapter}
\bauthor{\bsnm{Hemmati}, \binits{H.}}:
\bctitle{Advances in techniques for test prioritization}.
In: \bbtitle{Advances in Computers}
vol. \bseriesno{112},
pp. \bfpage{185}--\blpage{221}.
\bpublisher{Elsevier},
\blocation{Amsterdam, The Netherlands}
(\byear{2019})
\end{bchapter}
\endbibitem

\bibitem{hao2014unified}
\begin{barticle}
\bauthor{\bsnm{Hao}, \binits{D.}},
\bauthor{\bsnm{Zhang}, \binits{L.}},
\bauthor{\bsnm{Zhang}, \binits{L.}},
\bauthor{\bsnm{Rothermel}, \binits{G.}},
\bauthor{\bsnm{Mei}, \binits{H.}}:
\batitle{A unified test case prioritization approach}.
\bjtitle{ACM Transactions on Software Engineering and Methodology (TOSEM)}
\bvolume{24}(\bissue{2}),
\bfpage{1}--\blpage{31}
(\byear{2014})
\end{barticle}
\endbibitem

\bibitem{yoo2012regression}
\begin{barticle}
\bauthor{\bsnm{Yoo}, \binits{S.}},
\bauthor{\bsnm{Harman}, \binits{M.}}:
\batitle{Regression testing minimization, selection and prioritization: a
  survey}.
\bjtitle{Software testing, verification and reliability}
\bvolume{22}(\bissue{2}),
\bfpage{67}--\blpage{120}
(\byear{2012})
\end{barticle}
\endbibitem

\bibitem{hettiarachchi2016risk}
\begin{barticle}
\bauthor{\bsnm{Hettiarachchi}, \binits{C.}},
\bauthor{\bsnm{Do}, \binits{H.}},
\bauthor{\bsnm{Choi}, \binits{B.}}:
\batitle{Risk-based test case prioritization using a fuzzy expert system}.
\bjtitle{Information and Software Technology}
\bvolume{69},
\bfpage{1}--\blpage{15}
(\byear{2016})
\end{barticle}
\endbibitem

\bibitem{srikanth2016requirements}
\begin{barticle}
\bauthor{\bsnm{Srikanth}, \binits{H.}},
\bauthor{\bsnm{Hettiarachchi}, \binits{C.}},
\bauthor{\bsnm{Do}, \binits{H.}}:
\batitle{Requirements based test prioritization using risk factors: An
  industrial study}.
\bjtitle{Information and Software Technology}
\bvolume{69},
\bfpage{71}--\blpage{83}
(\byear{2016})
\end{barticle}
\endbibitem

\bibitem{salehie2011prioritizing}
\begin{bchapter}
\bauthor{\bsnm{Salehie}, \binits{M.}},
\bauthor{\bsnm{Li}, \binits{S.}},
\bauthor{\bsnm{Tahvildari}, \binits{L.}},
\bauthor{\bsnm{Dara}, \binits{R.}},
\bauthor{\bsnm{Li}, \binits{S.}},
\bauthor{\bsnm{Moore}, \binits{M.}}:
\bctitle{Prioritizing requirements-based regression test cases: A goal-driven
  practice}.
In: \bbtitle{2011 15th European Conference on Software Maintenance and
  Reengineering},
pp. \bfpage{329}--\blpage{332}
(\byear{2011}).
\bcomment{IEEE}
\end{bchapter}
\endbibitem

\bibitem{alves2016prioritizing}
\begin{barticle}
\bauthor{\bsnm{Alves}, \binits{E.L.}},
\bauthor{\bsnm{Machado}, \binits{P.D.}},
\bauthor{\bsnm{Massoni}, \binits{T.}},
\bauthor{\bsnm{Kim}, \binits{M.}}:
\batitle{Prioritizing test cases for early detection of refactoring faults}.
\bjtitle{Software Testing, Verification and Reliability}
\bvolume{26}(\bissue{5}),
\bfpage{402}--\blpage{426}
(\byear{2016})
\end{barticle}
\endbibitem

\bibitem{saha2015information}
\begin{bchapter}
\bauthor{\bsnm{Saha}, \binits{R.K.}},
\bauthor{\bsnm{Zhang}, \binits{L.}},
\bauthor{\bsnm{Khurshid}, \binits{S.}},
\bauthor{\bsnm{Perry}, \binits{D.E.}}:
\bctitle{An information retrieval approach for regression test prioritization
  based on program changes}.
In: \bbtitle{2015 IEEE/ACM 37th IEEE International Conference on Software
  Engineering},
vol. \bseriesno{1},
pp. \bfpage{268}--\blpage{279}
(\byear{2015}).
\bcomment{IEEE}
\end{bchapter}
\endbibitem

\bibitem{panda2016slice}
\begin{botherref}
\oauthor{\bsnm{Panda}, \binits{S.}},
\oauthor{\bsnm{Munjal}, \binits{D.}},
\oauthor{\bsnm{Mohapatra}, \binits{D.P.}}:
A slice-based change impact analysis for regression test case prioritization of
  object-oriented programs.
Advances in Software Engineering
\textbf{2016}
(2016)
\end{botherref}
\endbibitem

\bibitem{noor2015similarity}
\begin{bchapter}
\bauthor{\bsnm{Noor}, \binits{T.B.}},
\bauthor{\bsnm{Hemmati}, \binits{H.}}:
\bctitle{A similarity-based approach for test case prioritization using
  historical failure data}.
In: \bbtitle{2015 IEEE 26th International Symposium on Software Reliability
  Engineering (ISSRE)},
pp. \bfpage{58}--\blpage{68}
(\byear{2015}).
\bcomment{IEEE}
\end{bchapter}
\endbibitem

\bibitem{khalilian2012improved}
\begin{barticle}
\bauthor{\bsnm{Khalilian}, \binits{A.}},
\bauthor{\bsnm{Azgomi}, \binits{M.A.}},
\bauthor{\bsnm{Fazlalizadeh}, \binits{Y.}}:
\batitle{An improved method for test case prioritization by incorporating
  historical test case data}.
\bjtitle{Science of Computer Programming}
\bvolume{78}(\bissue{1}),
\bfpage{93}--\blpage{116}
(\byear{2012})
\end{barticle}
\endbibitem

\bibitem{rahman2018prioritizing}
\begin{barticle}
\bauthor{\bsnm{Rahman}, \binits{M.A.}},
\bauthor{\bsnm{Hasan}, \binits{M.A.}},
\bauthor{\bsnm{Siddik}, \binits{M.S.}}:
\batitle{Prioritizing dissimilar test cases in regression testing using
  historical failure data}.
\bjtitle{International Journal of Computer Applications}
\bvolume{975},
\bfpage{8887}
(\byear{2018})
\end{barticle}
\endbibitem

\bibitem{wang2017qtep}
\begin{bchapter}
\bauthor{\bsnm{Wang}, \binits{S.}},
\bauthor{\bsnm{Nam}, \binits{J.}},
\bauthor{\bsnm{Tan}, \binits{L.}}:
\bctitle{Qtep: quality-aware test case prioritization}.
In: \bbtitle{Proceedings of the 2017 11th Joint Meeting on Foundations of
  Software Engineering},
pp. \bfpage{523}--\blpage{534}
(\byear{2017})
\end{bchapter}
\endbibitem

\bibitem{paterson2019empirical}
\begin{bchapter}
\bauthor{\bsnm{Paterson}, \binits{D.}},
\bauthor{\bsnm{Campos}, \binits{J.}},
\bauthor{\bsnm{Abreu}, \binits{R.}},
\bauthor{\bsnm{Kapfhammer}, \binits{G.M.}},
\bauthor{\bsnm{Fraser}, \binits{G.}},
\bauthor{\bsnm{McMinn}, \binits{P.}}:
\bctitle{An empirical study on the use of defect prediction for test case
  prioritization}.
In: \bbtitle{2019 12th IEEE Conference on Software Testing, Validation and
  Verification (ICST)},
pp. \bfpage{346}--\blpage{357}
(\byear{2019}).
\bcomment{IEEE}
\end{bchapter}
\endbibitem

\bibitem{eghbali2019supervised}
\begin{bchapter}
\bauthor{\bsnm{Eghbali}, \binits{S.}},
\bauthor{\bsnm{Kudva}, \binits{V.}},
\bauthor{\bsnm{Rothermel}, \binits{G.}},
\bauthor{\bsnm{Tahvildari}, \binits{L.}}:
\bctitle{Supervised tie breaking in test case prioritization}.
In: \bbtitle{2019 IEEE/ACM 41st International Conference on Software
  Engineering: Companion Proceedings (ICSE-Companion)},
pp. \bfpage{242}--\blpage{243}
(\byear{2019}).
\bcomment{IEEE}
\end{bchapter}
\endbibitem

\bibitem{mahdieh2020incorporating}
\begin{barticle}
\bauthor{\bsnm{Mahdieh}, \binits{M.}},
\bauthor{\bsnm{Mirian-Hosseinabadi}, \binits{S.-H.}},
\bauthor{\bsnm{Etemadi}, \binits{K.}},
\bauthor{\bsnm{Nosrati}, \binits{A.}},
\bauthor{\bsnm{Jalali}, \binits{S.}}:
\batitle{Incorporating fault-proneness estimations into coverage-based test
  case prioritization methods}.
\bjtitle{Information and Software Technology}
\bvolume{121},
\bfpage{106269}
(\byear{2020})
\end{barticle}
\endbibitem

\bibitem{zimmermann2007predicting}
\begin{bchapter}
\bauthor{\bsnm{Zimmermann}, \binits{T.}},
\bauthor{\bsnm{Premraj}, \binits{R.}},
\bauthor{\bsnm{Zeller}, \binits{A.}}:
\bctitle{Predicting defects for eclipse}.
In: \bbtitle{Third International Workshop on Predictor Models in Software
  Engineering (PROMISE'07: ICSE Workshops 2007)},
pp. \bfpage{9}--\blpage{9}
(\byear{2007}).
\bcomment{IEEE}
\end{bchapter}
\endbibitem

\bibitem{ostrand2005predicting}
\begin{barticle}
\bauthor{\bsnm{Ostrand}, \binits{T.J.}},
\bauthor{\bsnm{Weyuker}, \binits{E.J.}},
\bauthor{\bsnm{Bell}, \binits{R.M.}}:
\batitle{Predicting the location and number of faults in large software
  systems}.
\bjtitle{IEEE Transactions on Software Engineering}
\bvolume{31}(\bissue{4}),
\bfpage{340}--\blpage{355}
(\byear{2005})
\end{barticle}
\endbibitem

\bibitem{menzies2006data}
\begin{barticle}
\bauthor{\bsnm{Menzies}, \binits{T.}},
\bauthor{\bsnm{Greenwald}, \binits{J.}},
\bauthor{\bsnm{Frank}, \binits{A.}}:
\batitle{Data mining static code attributes to learn defect predictors}.
\bjtitle{IEEE transactions on software engineering}
\bvolume{33}(\bissue{1}),
\bfpage{2}--\blpage{13}
(\byear{2006})
\end{barticle}
\endbibitem

\bibitem{leon2003comparison}
\begin{bchapter}
\bauthor{\bsnm{Leon}, \binits{D.}},
\bauthor{\bsnm{Podgurski}, \binits{A.}}:
\bctitle{A comparison of coverage-based and distribution-based techniques for
  filtering and prioritizing test cases}.
In: \bbtitle{14th International Symposium on Software Reliability Engineering,
  2003. ISSRE 2003.},
pp. \bfpage{442}--\blpage{453}
(\byear{2003}).
\bcomment{IEEE}
\end{bchapter}
\endbibitem

\bibitem{yoo2009clustering}
\begin{bchapter}
\bauthor{\bsnm{Yoo}, \binits{S.}},
\bauthor{\bsnm{Harman}, \binits{M.}},
\bauthor{\bsnm{Tonella}, \binits{P.}},
\bauthor{\bsnm{Susi}, \binits{A.}}:
\bctitle{Clustering test cases to achieve effective and scalable prioritisation
  incorporating expert knowledge}.
In: \bbtitle{Proceedings of the Eighteenth International Symposium on Software
  Testing and Analysis},
pp. \bfpage{201}--\blpage{212}
(\byear{2009})
\end{bchapter}
\endbibitem

\bibitem{jiang2009adaptive}
\begin{bchapter}
\bauthor{\bsnm{Jiang}, \binits{B.}},
\bauthor{\bsnm{Zhang}, \binits{Z.}},
\bauthor{\bsnm{Chan}, \binits{W.K.}},
\bauthor{\bsnm{Tse}, \binits{T.}}:
\bctitle{Adaptive random test case prioritization}.
In: \bbtitle{2009 IEEE/ACM International Conference on Automated Software
  Engineering},
pp. \bfpage{233}--\blpage{244}
(\byear{2009}).
\bcomment{IEEE}
\end{bchapter}
\endbibitem

\bibitem{fang2014similarity}
\begin{barticle}
\bauthor{\bsnm{Fang}, \binits{C.}},
\bauthor{\bsnm{Chen}, \binits{Z.}},
\bauthor{\bsnm{Wu}, \binits{K.}},
\bauthor{\bsnm{Zhao}, \binits{Z.}}:
\batitle{Similarity-based test case prioritization using ordered sequences of
  program entities}.
\bjtitle{Software Quality Journal}
\bvolume{22}(\bissue{2}),
\bfpage{335}--\blpage{361}
(\byear{2014})
\end{barticle}
\endbibitem

\bibitem{ledru2012prioritizing}
\begin{barticle}
\bauthor{\bsnm{Ledru}, \binits{Y.}},
\bauthor{\bsnm{Petrenko}, \binits{A.}},
\bauthor{\bsnm{Boroday}, \binits{S.}},
\bauthor{\bsnm{Mandran}, \binits{N.}}:
\batitle{Prioritizing test cases with string distances}.
\bjtitle{Automated Software Engineering}
\bvolume{19}(\bissue{1}),
\bfpage{65}--\blpage{95}
(\byear{2012})
\end{barticle}
\endbibitem

\bibitem{just2014defects4j}
\begin{bchapter}
\bauthor{\bsnm{Just}, \binits{R.}},
\bauthor{\bsnm{Jalali}, \binits{D.}},
\bauthor{\bsnm{Ernst}, \binits{M.D.}}:
\bctitle{Defects4j: A database of existing faults to enable controlled testing
  studies for java programs}.
In: \bbtitle{Proceedings of the 2014 International Symposium on Software
  Testing and Analysis},
pp. \bfpage{437}--\blpage{440}
(\byear{2014}).
\bcomment{ACM}
\end{bchapter}
\endbibitem

\bibitem{rothermel2002empirical}
\begin{barticle}
\bauthor{\bsnm{Rothermel}, \binits{G.}},
\bauthor{\bsnm{Harrold}, \binits{M.J.}},
\bauthor{\bsnm{Von~Ronne}, \binits{J.}},
\bauthor{\bsnm{Hong}, \binits{C.}}:
\batitle{Empirical studies of test-suite reduction}.
\bjtitle{Software Testing, Verification and Reliability}
\bvolume{12}(\bissue{4}),
\bfpage{219}--\blpage{249}
(\byear{2002})
\end{barticle}
\endbibitem

\bibitem{hao2015optimal}
\begin{barticle}
\bauthor{\bsnm{Hao}, \binits{D.}},
\bauthor{\bsnm{Zhang}, \binits{L.}},
\bauthor{\bsnm{Zang}, \binits{L.}},
\bauthor{\bsnm{Wang}, \binits{Y.}},
\bauthor{\bsnm{Wu}, \binits{X.}},
\bauthor{\bsnm{Xie}, \binits{T.}}:
\batitle{To be optimal or not in test-case prioritization}.
\bjtitle{IEEE Transactions on Software Engineering}
\bvolume{42}(\bissue{5}),
\bfpage{490}--\blpage{505}
(\byear{2015})
\end{barticle}
\endbibitem

\bibitem{usingz}
\begin{bbook}
\bauthor{\bsnm{Woodcock}, \binits{J.}},
\bauthor{\bsnm{Davies}, \binits{J.}}:
\bbtitle{Using Z: Specification, Refinement, and Proof}
vol. \bseriesno{39}.
\bpublisher{Prentice Hall},
\blocation{Englewood Cliffs, New Jersey, USA}
(\byear{1996})
\end{bbook}
\endbibitem

\bibitem{engstrom2011improving}
\begin{bchapter}
\bauthor{\bsnm{Engstr{\"o}m}, \binits{E.}},
\bauthor{\bsnm{Runeson}, \binits{P.}},
\bauthor{\bsnm{Ljung}, \binits{A.}}:
\bctitle{Improving regression testing transparency and efficiency with
  history-based prioritization--an industrial case study}.
In: \bbtitle{2011 Fourth IEEE International Conference on Software Testing,
  Verification and Validation},
pp. \bfpage{367}--\blpage{376}
(\byear{2011}).
\bcomment{IEEE}
\end{bchapter}
\endbibitem

\bibitem{catal2013test}
\begin{barticle}
\bauthor{\bsnm{Catal}, \binits{C.}},
\bauthor{\bsnm{Mishra}, \binits{D.}}:
\batitle{Test case prioritization: a systematic mapping study}.
\bjtitle{Software Quality Journal}
\bvolume{21}(\bissue{3}),
\bfpage{445}--\blpage{478}
(\byear{2013})
\end{barticle}
\endbibitem

\bibitem{mei2012static}
\begin{barticle}
\bauthor{\bsnm{Mei}, \binits{H.}},
\bauthor{\bsnm{Hao}, \binits{D.}},
\bauthor{\bsnm{Zhang}, \binits{L.}},
\bauthor{\bsnm{Zhang}, \binits{L.}},
\bauthor{\bsnm{Zhou}, \binits{J.}},
\bauthor{\bsnm{Rothermel}, \binits{G.}}:
\batitle{A static approach to prioritizing junit test cases}.
\bjtitle{Software Engineering, IEEE Transactions on}
\bvolume{38}(\bissue{6}),
\bfpage{1258}--\blpage{1275}
(\byear{2012})
\end{barticle}
\endbibitem

\bibitem{ashraf2012value}
\begin{bchapter}
\bauthor{\bsnm{Ashraf}, \binits{E.}},
\bauthor{\bsnm{Rauf}, \binits{A.}},
\bauthor{\bsnm{Mahmood}, \binits{K.}}:
\bctitle{Value based regression test case prioritization}.
In: \bbtitle{Proceedings of the World Congress on Engineering and Computer
  Science},
vol. \bseriesno{1},
pp. \bfpage{24}--\blpage{26}
(\byear{2012})
\end{bchapter}
\endbibitem

\bibitem{elbaum2004selecting}
\begin{barticle}
\bauthor{\bsnm{Elbaum}, \binits{S.}},
\bauthor{\bsnm{Rothermel}, \binits{G.}},
\bauthor{\bsnm{Kanduri}, \binits{S.}},
\bauthor{\bsnm{Malishevsky}, \binits{A.G.}}:
\batitle{Selecting a cost-effective test case prioritization technique}.
\bjtitle{Software Quality Journal}
\bvolume{12}(\bissue{3}),
\bfpage{185}--\blpage{210}
(\byear{2004})
\end{barticle}
\endbibitem

\bibitem{nam2014survey}
\begin{botherref}
\oauthor{\bsnm{Nam}, \binits{J.}}:
Survey on software defect prediction.
Department of Compter Science and Engineerning, The Hong Kong University of
  Science and Technology, Tech. Rep
(2014)
\end{botherref}
\endbibitem

\bibitem{mccabe1976complexity}
\begin{botherref}
\oauthor{\bsnm{McCabe}, \binits{T.J.}}:
A complexity measure.
IEEE Transactions on software Engineering
(4),
308--320
(1976)
\end{botherref}
\endbibitem

\bibitem{halstead1977elements}
\begin{bbook}
\bauthor{\bsnm{Halstead}, \binits{M.H.}}:
\bbtitle{Elements of Software Science (Operating and Programming Systems
  Series)}.
\bpublisher{Elsevier Science Inc.},
\blocation{Amsterdam, The Netherlands}
(\byear{1977})
\end{bbook}
\endbibitem

\bibitem{chidamber1994metrics}
\begin{barticle}
\bauthor{\bsnm{Chidamber}, \binits{S.R.}},
\bauthor{\bsnm{Kemerer}, \binits{C.F.}}:
\batitle{A metrics suite for object oriented design}.
\bjtitle{IEEE Transactions on software engineering}
\bvolume{20}(\bissue{6}),
\bfpage{476}--\blpage{493}
(\byear{1994})
\end{barticle}
\endbibitem

\bibitem{harrison1998evaluation}
\begin{barticle}
\bauthor{\bsnm{Harrison}, \binits{R.}},
\bauthor{\bsnm{Counsell}, \binits{S.J.}},
\bauthor{\bsnm{Nithi}, \binits{R.V.}}:
\batitle{An evaluation of the mood set of object-oriented software metrics}.
\bjtitle{IEEE Transactions on Software Engineering}
\bvolume{24}(\bissue{6}),
\bfpage{491}--\blpage{496}
(\byear{1998})
\end{barticle}
\endbibitem

\bibitem{bansiya2002hierarchical}
\begin{barticle}
\bauthor{\bsnm{Bansiya}, \binits{J.}},
\bauthor{\bsnm{Davis}, \binits{C.G.}}:
\batitle{A hierarchical model for object-oriented design quality assessment}.
\bjtitle{IEEE Transactions on software engineering}
\bvolume{28}(\bissue{1}),
\bfpage{4}--\blpage{17}
(\byear{2002})
\end{barticle}
\endbibitem

\bibitem{e1994candidate}
\begin{barticle}
\bauthor{\bparticle{e} \bsnm{Abreu}, \binits{F.B.}},
\bauthor{\bsnm{Carapu{\c{c}}a}, \binits{R.}}:
\batitle{Candidate metrics for object-oriented software within a taxonomy
  framework}.
\bjtitle{Journal of Systems and Software}
\bvolume{26}(\bissue{1}),
\bfpage{87}--\blpage{96}
(\byear{1994})
\end{barticle}
\endbibitem

\bibitem{menzies2010defect}
\begin{barticle}
\bauthor{\bsnm{Menzies}, \binits{T.}},
\bauthor{\bsnm{Milton}, \binits{Z.}},
\bauthor{\bsnm{Turhan}, \binits{B.}},
\bauthor{\bsnm{Cukic}, \binits{B.}},
\bauthor{\bsnm{Jiang}, \binits{Y.}},
\bauthor{\bsnm{Bener}, \binits{A.}}:
\batitle{Defect prediction from static code features: current results,
  limitations, new approaches}.
\bjtitle{Automated Software Engineering}
\bvolume{17}(\bissue{4}),
\bfpage{375}--\blpage{407}
(\byear{2010})
\end{barticle}
\endbibitem

\bibitem{menzies2007data}
\begin{barticle}
\bauthor{\bsnm{Menzies}, \binits{T.}},
\bauthor{\bsnm{Greenwald}, \binits{J.}},
\bauthor{\bsnm{Frank}, \binits{A.}}:
\batitle{Data mining static code attributes to learn defect predictors}.
\bjtitle{Software Engineering, IEEE Transactions on}
\bvolume{33}(\bissue{1}),
\bfpage{2}--\blpage{13}
(\byear{2007})
\end{barticle}
\endbibitem

\bibitem{klas2010transparent}
\begin{bchapter}
\bauthor{\bsnm{Kl{\"a}s}, \binits{M.}},
\bauthor{\bsnm{Elberzhager}, \binits{F.}},
\bauthor{\bsnm{M{\"u}nch}, \binits{J.}},
\bauthor{\bsnm{Hartjes}, \binits{K.}},
\bauthor{\bsnm{Von~Graevemeyer}, \binits{O.}}:
\bctitle{Transparent combination of expert and measurement data for defect
  prediction: an industrial case study}.
In: \bbtitle{Proceedings of the 32nd ACM/IEEE International Conference on
  Software Engineering-Volume 2},
pp. \bfpage{119}--\blpage{128}
(\byear{2010}).
\bcomment{ACM}
\end{bchapter}
\endbibitem

\bibitem{pinzger2008can}
\begin{bchapter}
\bauthor{\bsnm{Pinzger}, \binits{M.}},
\bauthor{\bsnm{Nagappan}, \binits{N.}},
\bauthor{\bsnm{Murphy}, \binits{B.}}:
\bctitle{Can developer-module networks predict failures?}
In: \bbtitle{Proceedings of the 16th ACM SIGSOFT International Symposium on
  Foundations of Software Engineering},
pp. \bfpage{2}--\blpage{12}
(\byear{2008}).
\bcomment{ACM}
\end{bchapter}
\endbibitem

\bibitem{meneely2008predicting}
\begin{bchapter}
\bauthor{\bsnm{Meneely}, \binits{A.}},
\bauthor{\bsnm{Williams}, \binits{L.}},
\bauthor{\bsnm{Snipes}, \binits{W.}},
\bauthor{\bsnm{Osborne}, \binits{J.}}:
\bctitle{Predicting failures with developer networks and social network
  analysis}.
In: \bbtitle{Proceedings of the 16th ACM SIGSOFT International Symposium on
  Foundations of Software Engineering},
pp. \bfpage{13}--\blpage{23}
(\byear{2008}).
\bcomment{ACM}
\end{bchapter}
\endbibitem

\bibitem{moser2008comparative}
\begin{bchapter}
\bauthor{\bsnm{Moser}, \binits{R.}},
\bauthor{\bsnm{Pedrycz}, \binits{W.}},
\bauthor{\bsnm{Succi}, \binits{G.}}:
\bctitle{A comparative analysis of the efficiency of change metrics and static
  code attributes for defect prediction}.
In: \bbtitle{Software Engineering, 2008. ICSE'08. ACM/IEEE 30th International
  Conference On},
pp. \bfpage{181}--\blpage{190}
(\byear{2008}).
\bcomment{IEEE}
\end{bchapter}
\endbibitem

\bibitem{weyuker2008too}
\begin{barticle}
\bauthor{\bsnm{Weyuker}, \binits{E.J.}},
\bauthor{\bsnm{Ostrand}, \binits{T.J.}},
\bauthor{\bsnm{Bell}, \binits{R.M.}}:
\batitle{Do too many cooks spoil the broth? using the number of developers to
  enhance defect prediction models}.
\bjtitle{Empirical Software Engineering}
\bvolume{13}(\bissue{5}),
\bfpage{539}--\blpage{559}
(\byear{2008})
\end{barticle}
\endbibitem

\bibitem{graves2000predicting}
\begin{barticle}
\bauthor{\bsnm{Graves}, \binits{T.L.}},
\bauthor{\bsnm{Karr}, \binits{A.F.}},
\bauthor{\bsnm{Marron}, \binits{J.S.}},
\bauthor{\bsnm{Siy}, \binits{H.}}:
\batitle{Predicting fault incidence using software change history}.
\bjtitle{IEEE Transactions on software engineering}
\bvolume{26}(\bissue{7}),
\bfpage{653}--\blpage{661}
(\byear{2000})
\end{barticle}
\endbibitem

\bibitem{li2018progress}
\begin{barticle}
\bauthor{\bsnm{Li}, \binits{Z.}},
\bauthor{\bsnm{Jing}, \binits{X.-Y.}},
\bauthor{\bsnm{Zhu}, \binits{X.}}:
\batitle{Progress on approaches to software defect prediction}.
\bjtitle{Iet Software}
\bvolume{12}(\bissue{3}),
\bfpage{161}--\blpage{175}
(\byear{2018})
\end{barticle}
\endbibitem

\bibitem{kanmani2007object}
\begin{barticle}
\bauthor{\bsnm{Kanmani}, \binits{S.}},
\bauthor{\bsnm{Uthariaraj}, \binits{V.R.}},
\bauthor{\bsnm{Sankaranarayanan}, \binits{V.}},
\bauthor{\bsnm{Thambidurai}, \binits{P.}}:
\batitle{Object-oriented software fault prediction using neural networks}.
\bjtitle{Information and software technology}
\bvolume{49}(\bissue{5}),
\bfpage{483}--\blpage{492}
(\byear{2007})
\end{barticle}
\endbibitem

\bibitem{elish2008predicting}
\begin{barticle}
\bauthor{\bsnm{Elish}, \binits{K.O.}},
\bauthor{\bsnm{Elish}, \binits{M.O.}}:
\batitle{Predicting defect-prone software modules using support vector
  machines}.
\bjtitle{Journal of Systems and Software}
\bvolume{81}(\bissue{5}),
\bfpage{649}--\blpage{660}
(\byear{2008})
\end{barticle}
\endbibitem

\bibitem{shivaji2009reducing}
\begin{bchapter}
\bauthor{\bsnm{Shivaji}, \binits{S.}},
\bauthor{\bsnm{Whitehead}, \binits{E.J.}},
\bauthor{\bsnm{Akella}, \binits{R.}},
\bauthor{\bsnm{Kim}, \binits{S.}}:
\bctitle{Reducing features to improve bug prediction}.
In: \bbtitle{2009 IEEE/ACM International Conference on Automated Software
  Engineering},
pp. \bfpage{600}--\blpage{604}
(\byear{2009}).
\bcomment{IEEE}
\end{bchapter}
\endbibitem

\bibitem{okutan2014software}
\begin{barticle}
\bauthor{\bsnm{Okutan}, \binits{A.}},
\bauthor{\bsnm{Y{\i}ld{\i}z}, \binits{O.T.}}:
\batitle{Software defect prediction using bayesian networks}.
\bjtitle{Empirical Software Engineering}
\bvolume{19}(\bissue{1}),
\bfpage{154}--\blpage{181}
(\byear{2014})
\end{barticle}
\endbibitem

\bibitem{jing2014dictionary}
\begin{bchapter}
\bauthor{\bsnm{Jing}, \binits{X.-Y.}},
\bauthor{\bsnm{Ying}, \binits{S.}},
\bauthor{\bsnm{Zhang}, \binits{Z.-W.}},
\bauthor{\bsnm{Wu}, \binits{S.-S.}},
\bauthor{\bsnm{Liu}, \binits{J.}}:
\bctitle{Dictionary learning based software defect prediction}.
In: \bbtitle{Proceedings of the 36th International Conference on Software
  Engineering},
pp. \bfpage{414}--\blpage{423}
(\byear{2014})
\end{bchapter}
\endbibitem

\bibitem{petric2016building}
\begin{bchapter}
\bauthor{\bsnm{Petri{\'c}}, \binits{J.}},
\bauthor{\bsnm{Bowes}, \binits{D.}},
\bauthor{\bsnm{Hall}, \binits{T.}},
\bauthor{\bsnm{Christianson}, \binits{B.}},
\bauthor{\bsnm{Baddoo}, \binits{N.}}:
\bctitle{Building an ensemble for software defect prediction based on diversity
  selection}.
In: \bbtitle{Proceedings of the 10th ACM/IEEE International Symposium on
  Empirical Software Engineering and Measurement},
pp. \bfpage{1}--\blpage{10}
(\byear{2016})
\end{bchapter}
\endbibitem

\bibitem{aljamaan2020software}
\begin{bchapter}
\bauthor{\bsnm{Aljamaan}, \binits{H.}},
\bauthor{\bsnm{Alazba}, \binits{A.}}:
\bctitle{Software defect prediction using tree-based ensembles}.
In: \bbtitle{Proceedings of the 16th ACM International Conference on Predictive
  Models and Data Analytics in Software Engineering},
pp. \bfpage{1}--\blpage{10}
(\byear{2020})
\end{bchapter}
\endbibitem

\bibitem{matloob2021software}
\begin{botherref}
\oauthor{\bsnm{Matloob}, \binits{F.}},
\oauthor{\bsnm{Ghazal}, \binits{T.M.}},
\oauthor{\bsnm{Taleb}, \binits{N.}},
\oauthor{\bsnm{Aftab}, \binits{S.}},
\oauthor{\bsnm{Ahmad}, \binits{M.}},
\oauthor{\bsnm{Khan}, \binits{M.A.}},
\oauthor{\bsnm{Abbas}, \binits{S.}},
\oauthor{\bsnm{Soomro}, \binits{T.R.}}:
Software defect prediction using ensemble learning: A systematic literature
  review.
IEEE Access
(2021)
\end{botherref}
\endbibitem

\bibitem{li2019software}
\begin{bchapter}
\bauthor{\bsnm{Li}, \binits{R.}},
\bauthor{\bsnm{Zhou}, \binits{L.}},
\bauthor{\bsnm{Zhang}, \binits{S.}},
\bauthor{\bsnm{Liu}, \binits{H.}},
\bauthor{\bsnm{Huang}, \binits{X.}},
\bauthor{\bsnm{Sun}, \binits{Z.}}:
\bctitle{Software defect prediction based on ensemble learning}.
In: \bbtitle{Proceedings of the 2019 2nd International Conference on Data
  Science and Information Technology},
pp. \bfpage{1}--\blpage{6}
(\byear{2019})
\end{bchapter}
\endbibitem

\bibitem{li2020systematic}
\begin{barticle}
\bauthor{\bsnm{Li}, \binits{N.}},
\bauthor{\bsnm{Shepperd}, \binits{M.}},
\bauthor{\bsnm{Guo}, \binits{Y.}}:
\batitle{A systematic review of unsupervised learning techniques for software
  defect prediction}.
\bjtitle{Information and Software Technology}
\bvolume{122},
\bfpage{106287}
(\byear{2020})
\end{barticle}
\endbibitem

\bibitem{nam2015clami}
\begin{bchapter}
\bauthor{\bsnm{Nam}, \binits{J.}},
\bauthor{\bsnm{Kim}, \binits{S.}}:
\bctitle{Clami: Defect prediction on unlabeled datasets}.
In: \bbtitle{2015 30th IEEE/ACM International Conference on Automated Software
  Engineering (ASE)},
pp. \bfpage{452}--\blpage{463}
(\byear{2015}).
\bcomment{IEEE}
\end{bchapter}
\endbibitem

\bibitem{bishnu2011software}
\begin{barticle}
\bauthor{\bsnm{Bishnu}, \binits{P.S.}},
\bauthor{\bsnm{Bhattacherjee}, \binits{V.}}:
\batitle{Software fault prediction using quad tree-based k-means clustering
  algorithm}.
\bjtitle{IEEE Transactions on knowledge and data engineering}
\bvolume{24}(\bissue{6}),
\bfpage{1146}--\blpage{1150}
(\byear{2011})
\end{barticle}
\endbibitem

\bibitem{zhang2016cross}
\begin{bchapter}
\bauthor{\bsnm{Zhang}, \binits{F.}},
\bauthor{\bsnm{Zheng}, \binits{Q.}},
\bauthor{\bsnm{Zou}, \binits{Y.}},
\bauthor{\bsnm{Hassan}, \binits{A.E.}}:
\bctitle{Cross-project defect prediction using a connectivity-based
  unsupervised classifier}.
In: \bbtitle{2016 IEEE/ACM 38th International Conference on Software
  Engineering (ICSE)},
pp. \bfpage{309}--\blpage{320}
(\byear{2016}).
\bcomment{IEEE}
\end{bchapter}
\endbibitem

\bibitem{yang2016effort}
\begin{bchapter}
\bauthor{\bsnm{Yang}, \binits{Y.}},
\bauthor{\bsnm{Zhou}, \binits{Y.}},
\bauthor{\bsnm{Liu}, \binits{J.}},
\bauthor{\bsnm{Zhao}, \binits{Y.}},
\bauthor{\bsnm{Lu}, \binits{H.}},
\bauthor{\bsnm{Xu}, \binits{L.}},
\bauthor{\bsnm{Xu}, \binits{B.}},
\bauthor{\bsnm{Leung}, \binits{H.}}:
\bctitle{Effort-aware just-in-time defect prediction: simple unsupervised
  models could be better than supervised models}.
In: \bbtitle{Proceedings of the 2016 24th ACM SIGSOFT International Symposium
  on Foundations of Software Engineering},
pp. \bfpage{157}--\blpage{168}
(\byear{2016})
\end{bchapter}
\endbibitem

\bibitem{fu2017revisiting}
\begin{bchapter}
\bauthor{\bsnm{Fu}, \binits{W.}},
\bauthor{\bsnm{Menzies}, \binits{T.}}:
\bctitle{Revisiting unsupervised learning for defect prediction}.
In: \bbtitle{Proceedings of the 2017 11th Joint Meeting on Foundations of
  Software Engineering},
pp. \bfpage{72}--\blpage{83}
(\byear{2017})
\end{bchapter}
\endbibitem

\bibitem{yan2017file}
\begin{bchapter}
\bauthor{\bsnm{Yan}, \binits{M.}},
\bauthor{\bsnm{Fang}, \binits{Y.}},
\bauthor{\bsnm{Lo}, \binits{D.}},
\bauthor{\bsnm{Xia}, \binits{X.}},
\bauthor{\bsnm{Zhang}, \binits{X.}}:
\bctitle{File-level defect prediction: Unsupervised vs. supervised models}.
In: \bbtitle{2017 ACM/IEEE International Symposium on Empirical Software
  Engineering and Measurement (ESEM)},
pp. \bfpage{344}--\blpage{353}
(\byear{2017}).
\bcomment{IEEE}
\end{bchapter}
\endbibitem

\bibitem{boucher2018software}
\begin{barticle}
\bauthor{\bsnm{Boucher}, \binits{A.}},
\bauthor{\bsnm{Badri}, \binits{M.}}:
\batitle{Software metrics thresholds calculation techniques to predict
  fault-proneness: An empirical comparison}.
\bjtitle{Information and Software Technology}
\bvolume{96},
\bfpage{38}--\blpage{67}
(\byear{2018})
\end{barticle}
\endbibitem

\bibitem{wang2016non}
\begin{barticle}
\bauthor{\bsnm{Wang}, \binits{T.}},
\bauthor{\bsnm{Zhang}, \binits{Z.}},
\bauthor{\bsnm{Jing}, \binits{X.}},
\bauthor{\bsnm{Liu}, \binits{Y.}}:
\batitle{Non-negative sparse-based semiboost for software defect prediction}.
\bjtitle{Software Testing, Verification and Reliability}
\bvolume{26}(\bissue{7}),
\bfpage{498}--\blpage{515}
(\byear{2016})
\end{barticle}
\endbibitem

\bibitem{zhang2017label}
\begin{barticle}
\bauthor{\bsnm{Zhang}, \binits{Z.-W.}},
\bauthor{\bsnm{Jing}, \binits{X.-Y.}},
\bauthor{\bsnm{Wang}, \binits{T.-J.}}:
\batitle{Label propagation based semi-supervised learning for software defect
  prediction}.
\bjtitle{Automated Software Engineering}
\bvolume{24}(\bissue{1}),
\bfpage{47}--\blpage{69}
(\byear{2017})
\end{barticle}
\endbibitem

\bibitem{kamei2012large}
\begin{barticle}
\bauthor{\bsnm{Kamei}, \binits{Y.}},
\bauthor{\bsnm{Shihab}, \binits{E.}},
\bauthor{\bsnm{Adams}, \binits{B.}},
\bauthor{\bsnm{Hassan}, \binits{A.E.}},
\bauthor{\bsnm{Mockus}, \binits{A.}},
\bauthor{\bsnm{Sinha}, \binits{A.}},
\bauthor{\bsnm{Ubayashi}, \binits{N.}}:
\batitle{A large-scale empirical study of just-in-time quality assurance}.
\bjtitle{IEEE Transactions on Software Engineering}
\bvolume{39}(\bissue{6}),
\bfpage{757}--\blpage{773}
(\byear{2012})
\end{barticle}
\endbibitem

\bibitem{yang2015deep}
\begin{bchapter}
\bauthor{\bsnm{Yang}, \binits{X.}},
\bauthor{\bsnm{Lo}, \binits{D.}},
\bauthor{\bsnm{Xia}, \binits{X.}},
\bauthor{\bsnm{Zhang}, \binits{Y.}},
\bauthor{\bsnm{Sun}, \binits{J.}}:
\bctitle{Deep learning for just-in-time defect prediction}.
In: \bbtitle{2015 IEEE International Conference on Software Quality,
  Reliability and Security},
pp. \bfpage{17}--\blpage{26}
(\byear{2015}).
\bcomment{IEEE}
\end{bchapter}
\endbibitem

\bibitem{wang2016automatically}
\begin{bchapter}
\bauthor{\bsnm{Wang}, \binits{S.}},
\bauthor{\bsnm{Liu}, \binits{T.}},
\bauthor{\bsnm{Tan}, \binits{L.}}:
\bctitle{Automatically learning semantic features for defect prediction}.
In: \bbtitle{2016 IEEE/ACM 38th International Conference on Software
  Engineering (ICSE)},
pp. \bfpage{297}--\blpage{308}
(\byear{2016}).
\bcomment{IEEE}
\end{bchapter}
\endbibitem

\bibitem{wang2018deep}
\begin{barticle}
\bauthor{\bsnm{Wang}, \binits{S.}},
\bauthor{\bsnm{Liu}, \binits{T.}},
\bauthor{\bsnm{Nam}, \binits{J.}},
\bauthor{\bsnm{Tan}, \binits{L.}}:
\batitle{Deep semantic feature learning for software defect prediction}.
\bjtitle{IEEE Transactions on Software Engineering}
\bvolume{46}(\bissue{12}),
\bfpage{1267}--\blpage{1293}
(\byear{2018})
\end{barticle}
\endbibitem

\bibitem{hoang2019deepjit}
\begin{bchapter}
\bauthor{\bsnm{Hoang}, \binits{T.}},
\bauthor{\bsnm{Dam}, \binits{H.K.}},
\bauthor{\bsnm{Kamei}, \binits{Y.}},
\bauthor{\bsnm{Lo}, \binits{D.}},
\bauthor{\bsnm{Ubayashi}, \binits{N.}}:
\bctitle{Deepjit: an end-to-end deep learning framework for just-in-time defect
  prediction}.
In: \bbtitle{2019 IEEE/ACM 16th International Conference on Mining Software
  Repositories (MSR)},
pp. \bfpage{34}--\blpage{45}
(\byear{2019}).
\bcomment{IEEE}
\end{bchapter}
\endbibitem

\bibitem{hoang2020cc2vec}
\begin{bchapter}
\bauthor{\bsnm{Hoang}, \binits{T.}},
\bauthor{\bsnm{Kang}, \binits{H.J.}},
\bauthor{\bsnm{Lo}, \binits{D.}},
\bauthor{\bsnm{Lawall}, \binits{J.}}:
\bctitle{Cc2vec: Distributed representations of code changes}.
In: \bbtitle{Proceedings of the ACM/IEEE 42nd International Conference on
  Software Engineering},
pp. \bfpage{518}--\blpage{529}
(\byear{2020})
\end{bchapter}
\endbibitem

\bibitem{pandey2020bpdet}
\begin{barticle}
\bauthor{\bsnm{Pandey}, \binits{S.K.}},
\bauthor{\bsnm{Mishra}, \binits{R.B.}},
\bauthor{\bsnm{Tripathi}, \binits{A.K.}}:
\batitle{Bpdet: An effective software bug prediction model using deep
  representation and ensemble learning techniques}.
\bjtitle{Expert Systems with Applications}
\bvolume{144},
\bfpage{113085}
(\byear{2020})
\end{barticle}
\endbibitem

\bibitem{majd2020sldeep}
\begin{barticle}
\bauthor{\bsnm{Majd}, \binits{A.}},
\bauthor{\bsnm{Vahidi-Asl}, \binits{M.}},
\bauthor{\bsnm{Khalilian}, \binits{A.}},
\bauthor{\bsnm{Poorsarvi-Tehrani}, \binits{P.}},
\bauthor{\bsnm{Haghighi}, \binits{H.}}:
\batitle{Sldeep: Statement-level software defect prediction using deep-learning
  model on static code features}.
\bjtitle{Expert Systems with Applications}
\bvolume{147},
\bfpage{113156}
(\byear{2020})
\end{barticle}
\endbibitem

\bibitem{deng2020software}
\begin{barticle}
\bauthor{\bsnm{Deng}, \binits{J.}},
\bauthor{\bsnm{Lu}, \binits{L.}},
\bauthor{\bsnm{Qiu}, \binits{S.}}:
\batitle{Software defect prediction via lstm}.
\bjtitle{IET Software}
\bvolume{14}(\bissue{4}),
\bfpage{443}--\blpage{450}
(\byear{2020})
\end{barticle}
\endbibitem

\bibitem{liang2019seml}
\begin{barticle}
\bauthor{\bsnm{Liang}, \binits{H.}},
\bauthor{\bsnm{Yu}, \binits{Y.}},
\bauthor{\bsnm{Jiang}, \binits{L.}},
\bauthor{\bsnm{Xie}, \binits{Z.}}:
\batitle{Seml: A semantic lstm model for software defect prediction}.
\bjtitle{IEEE Access}
\bvolume{7},
\bfpage{83812}--\blpage{83824}
(\byear{2019})
\end{barticle}
\endbibitem

\bibitem{tong2018software}
\begin{barticle}
\bauthor{\bsnm{Tong}, \binits{H.}},
\bauthor{\bsnm{Liu}, \binits{B.}},
\bauthor{\bsnm{Wang}, \binits{S.}}:
\batitle{Software defect prediction using stacked denoising autoencoders and
  two-stage ensemble learning}.
\bjtitle{Information and Software Technology}
\bvolume{96},
\bfpage{94}--\blpage{111}
(\byear{2018})
\end{barticle}
\endbibitem

\bibitem{zhu2020within}
\begin{barticle}
\bauthor{\bsnm{Zhu}, \binits{K.}},
\bauthor{\bsnm{Zhang}, \binits{N.}},
\bauthor{\bsnm{Ying}, \binits{S.}},
\bauthor{\bsnm{Zhu}, \binits{D.}}:
\batitle{Within-project and cross-project just-in-time defect prediction based
  on denoising autoencoder and convolutional neural network}.
\bjtitle{IET Software}
\bvolume{14}(\bissue{3}),
\bfpage{185}--\blpage{195}
(\byear{2020})
\end{barticle}
\endbibitem

\bibitem{li2017software}
\begin{bchapter}
\bauthor{\bsnm{Li}, \binits{J.}},
\bauthor{\bsnm{He}, \binits{P.}},
\bauthor{\bsnm{Zhu}, \binits{J.}},
\bauthor{\bsnm{Lyu}, \binits{M.R.}}:
\bctitle{Software defect prediction via convolutional neural network}.
In: \bbtitle{2017 IEEE International Conference on Software Quality,
  Reliability and Security (QRS)},
pp. \bfpage{318}--\blpage{328}
(\byear{2017}).
\bcomment{IEEE}
\end{bchapter}
\endbibitem

\bibitem{xu2019ldfr}
\begin{barticle}
\bauthor{\bsnm{Xu}, \binits{Z.}},
\bauthor{\bsnm{Li}, \binits{S.}},
\bauthor{\bsnm{Xu}, \binits{J.}},
\bauthor{\bsnm{Liu}, \binits{J.}},
\bauthor{\bsnm{Luo}, \binits{X.}},
\bauthor{\bsnm{Zhang}, \binits{Y.}},
\bauthor{\bsnm{Zhang}, \binits{T.}},
\bauthor{\bsnm{Keung}, \binits{J.}},
\bauthor{\bsnm{Tang}, \binits{Y.}}:
\batitle{Ldfr: Learning deep feature representation for software defect
  prediction}.
\bjtitle{Journal of Systems and Software}
\bvolume{158},
\bfpage{110402}
(\byear{2019})
\end{barticle}
\endbibitem

\bibitem{yedida2021value}
\begin{botherref}
\oauthor{\bsnm{Yedida}, \binits{R.}},
\oauthor{\bsnm{Menzies}, \binits{T.}}:
On the value of oversampling for deep learning in software defect prediction.
IEEE Transactions on Software Engineering
(2021)
\end{botherref}
\endbibitem

\bibitem{mathur2010foundations}
\begin{botherref}
\oauthor{\bsnm{Mathur}, \binits{A.P.}}:
Foundations of software testing.
Addison-Wesley Professional 11. Academic Accommodation Policy
(2010)
\end{botherref}
\endbibitem

\bibitem{kandil2017cluster}
\begin{barticle}
\bauthor{\bsnm{Kandil}, \binits{P.}},
\bauthor{\bsnm{Moussa}, \binits{S.}},
\bauthor{\bsnm{Badr}, \binits{N.}}:
\batitle{Cluster-based test cases prioritization and selection technique for
  agile regression testing}.
\bjtitle{Journal of Software: Evolution and Process}
\bvolume{29}(\bissue{6}),
\bfpage{1794}
(\byear{2017})
\end{barticle}
\endbibitem

\bibitem{pei131dynamic}
\begin{botherref}
\oauthor{\bsnm{Pei}, \binits{H.}},
\oauthor{\bsnm{Yin}, \binits{B.}},
\oauthor{\bsnm{Xie}, \binits{M.}},
\oauthor{\bsnm{Cai}, \binits{K.-Y.}}:
Dynamic random testing with test case clustering and distance-based parameter
  adjustment.
Information and Software Technology
\textbf{131},
106470
\end{botherref}
\endbibitem

\bibitem{fu2017coverage}
\begin{barticle}
\bauthor{\bsnm{Fu}, \binits{W.}},
\bauthor{\bsnm{Yu}, \binits{H.}},
\bauthor{\bsnm{Fan}, \binits{G.}},
\bauthor{\bsnm{Ji}, \binits{X.}}:
\batitle{Coverage-based clustering and scheduling approach for test case
  prioritization}.
\bjtitle{IEICE TRANSACTIONS on Information and Systems}
\bvolume{100}(\bissue{6}),
\bfpage{1218}--\blpage{1230}
(\byear{2017})
\end{barticle}
\endbibitem

\bibitem{shrivathsan2019novel}
\begin{barticle}
\bauthor{\bsnm{Shrivathsan}, \binits{A.}},
\bauthor{\bsnm{Ravichandran}, \binits{K.}},
\bauthor{\bsnm{Krishankumar}, \binits{R.}},
\bauthor{\bsnm{Sangeetha}, \binits{V.}},
\bauthor{\bsnm{Kar}, \binits{S.}},
\bauthor{\bsnm{Ziemba}, \binits{P.}},
\bauthor{\bsnm{Jankowski}, \binits{J.}}:
\batitle{Novel fuzzy clustering methods for test case prioritization in
  software projects}.
\bjtitle{Symmetry}
\bvolume{11}(\bissue{11}),
\bfpage{1400}
(\byear{2019})
\end{barticle}
\endbibitem

\bibitem{chen2018test}
\begin{barticle}
\bauthor{\bsnm{Chen}, \binits{J.}},
\bauthor{\bsnm{Zhu}, \binits{L.}},
\bauthor{\bsnm{Chen}, \binits{T.Y.}},
\bauthor{\bsnm{Towey}, \binits{D.}},
\bauthor{\bsnm{Kuo}, \binits{F.-C.}},
\bauthor{\bsnm{Huang}, \binits{R.}},
\bauthor{\bsnm{Guo}, \binits{Y.}}:
\batitle{Test case prioritization for object-oriented software: An adaptive
  random sequence approach based on clustering}.
\bjtitle{Journal of Systems and Software}
\bvolume{135},
\bfpage{107}--\blpage{125}
(\byear{2018})
\end{barticle}
\endbibitem

\bibitem{xu2015comprehensive}
\begin{barticle}
\bauthor{\bsnm{Xu}, \binits{D.}},
\bauthor{\bsnm{Tian}, \binits{Y.}}:
\batitle{A comprehensive survey of clustering algorithms}.
\bjtitle{Annals of Data Science}
\bvolume{2}(\bissue{2}),
\bfpage{165}--\blpage{193}
(\byear{2015})
\end{barticle}
\endbibitem

\bibitem{d2012evaluating}
\begin{barticle}
\bauthor{\bsnm{D’Ambros}, \binits{M.}},
\bauthor{\bsnm{Lanza}, \binits{M.}},
\bauthor{\bsnm{Robbes}, \binits{R.}}:
\batitle{Evaluating defect prediction approaches: a benchmark and an extensive
  comparison}.
\bjtitle{Empirical Software Engineering}
\bvolume{17}(\bissue{4}),
\bfpage{531}--\blpage{577}
(\byear{2012})
\end{barticle}
\endbibitem

\bibitem{lessmann2008benchmarking}
\begin{barticle}
\bauthor{\bsnm{Lessmann}, \binits{S.}},
\bauthor{\bsnm{Baesens}, \binits{B.}},
\bauthor{\bsnm{Mues}, \binits{C.}},
\bauthor{\bsnm{Pietsch}, \binits{S.}}:
\batitle{Benchmarking classification models for software defect prediction: A
  proposed framework and novel findings}.
\bjtitle{Software Engineering, IEEE Transactions on}
\bvolume{34}(\bissue{4}),
\bfpage{485}--\blpage{496}
(\byear{2008})
\end{barticle}
\endbibitem

\bibitem{guo2004robust}
\begin{bchapter}
\bauthor{\bsnm{Guo}, \binits{L.}},
\bauthor{\bsnm{Ma}, \binits{Y.}},
\bauthor{\bsnm{Cukic}, \binits{B.}},
\bauthor{\bsnm{Singh}, \binits{H.}}:
\bctitle{Robust prediction of fault-proneness by random forests}.
In: \bbtitle{15th International Symposium on Software Reliability Engineering},
pp. \bfpage{417}--\blpage{428}
(\byear{2004}).
\bcomment{IEEE}
\end{bchapter}
\endbibitem

\bibitem{ghotra2015revisiting}
\begin{bchapter}
\bauthor{\bsnm{Ghotra}, \binits{B.}},
\bauthor{\bsnm{McIntosh}, \binits{S.}},
\bauthor{\bsnm{Hassan}, \binits{A.E.}}:
\bctitle{Revisiting the impact of classification techniques on the performance
  of defect prediction models}.
In: \bbtitle{2015 IEEE/ACM 37th IEEE International Conference on Software
  Engineering},
vol. \bseriesno{1},
pp. \bfpage{789}--\blpage{800}
(\byear{2015}).
\bcomment{IEEE}
\end{bchapter}
\endbibitem

\bibitem{hastie2008elements}
\begin{botherref}
\oauthor{\bsnm{Hastie}, \binits{T.}},
\oauthor{\bsnm{Tibshirani}, \binits{R.}},
\oauthor{\bsnm{Friedman}, \binits{J.}}:
The elements of statistical learning, 2nd edn Berlin.
Germany: Springer.[Google Scholar]
(2008)
\end{botherref}
\endbibitem

\bibitem{khoshgoftaar2010attribute}
\begin{bchapter}
\bauthor{\bsnm{Khoshgoftaar}, \binits{T.M.}},
\bauthor{\bsnm{Gao}, \binits{K.}},
\bauthor{\bsnm{Seliya}, \binits{N.}}:
\bctitle{Attribute selection and imbalanced data: Problems in software defect
  prediction}.
In: \bbtitle{2010 22nd IEEE International Conference on Tools with Artificial
  Intelligence},
vol. \bseriesno{1},
pp. \bfpage{137}--\blpage{144}
(\byear{2010}).
\bcomment{IEEE}
\end{bchapter}
\endbibitem

\bibitem{song2018comprehensive}
\begin{barticle}
\bauthor{\bsnm{Song}, \binits{Q.}},
\bauthor{\bsnm{Guo}, \binits{Y.}},
\bauthor{\bsnm{Shepperd}, \binits{M.}}:
\batitle{A comprehensive investigation of the role of imbalanced learning for
  software defect prediction}.
\bjtitle{IEEE Transactions on Software Engineering}
\bvolume{45}(\bissue{12}),
\bfpage{1253}--\blpage{1269}
(\byear{2018})
\end{barticle}
\endbibitem

\bibitem{chawla2002smote}
\begin{barticle}
\bauthor{\bsnm{Chawla}, \binits{N.V.}},
\bauthor{\bsnm{Bowyer}, \binits{K.W.}},
\bauthor{\bsnm{Hall}, \binits{L.O.}},
\bauthor{\bsnm{Kegelmeyer}, \binits{W.P.}}:
\batitle{Smote: synthetic minority over-sampling technique}.
\bjtitle{Journal of artificial intelligence research}
\bvolume{16},
\bfpage{321}--\blpage{357}
(\byear{2002})
\end{barticle}
\endbibitem

\bibitem{carlson2011clustering}
\begin{bchapter}
\bauthor{\bsnm{Carlson}, \binits{R.}},
\bauthor{\bsnm{Do}, \binits{H.}},
\bauthor{\bsnm{Denton}, \binits{A.}}:
\bctitle{A clustering approach to improving test case prioritization: An
  industrial case study.}
In: \bbtitle{ICSM},
vol. \bseriesno{11},
pp. \bfpage{382}--\blpage{391}
(\byear{2011})
\end{bchapter}
\endbibitem

\bibitem{palma2018improvement}
\begin{bchapter}
\bauthor{\bsnm{Palma}, \binits{F.}},
\bauthor{\bsnm{Abdou}, \binits{T.}},
\bauthor{\bsnm{Bener}, \binits{A.}},
\bauthor{\bsnm{Maidens}, \binits{J.}},
\bauthor{\bsnm{Liu}, \binits{S.}}:
\bctitle{An improvement to test case failure prediction in the context of test
  case prioritization}.
In: \bbtitle{Proceedings of the 14th International Conference on Predictive
  Models and Data Analytics in Software Engineering},
pp. \bfpage{80}--\blpage{89}
(\byear{2018})
\end{bchapter}
\endbibitem

\bibitem{abou2021detrimental}
\begin{barticle}
\bauthor{\bsnm{Abou~Assi}, \binits{R.}},
\bauthor{\bsnm{Masri}, \binits{W.}},
\bauthor{\bsnm{Trad}, \binits{C.}}:
\batitle{How detrimental is coincidental correctness to coverage-based fault
  detection and localization? an empirical study}.
\bjtitle{Software Testing, Verification and Reliability}
\bvolume{31}(\bissue{5}),
\bfpage{1762}
(\byear{2021})
\end{barticle}
\endbibitem

\bibitem{yao2020assessing}
\begin{bchapter}
\bauthor{\bsnm{Yao}, \binits{J.}},
\bauthor{\bsnm{Shepperd}, \binits{M.}}:
\bctitle{Assessing software defection prediction performance: Why using the
  matthews correlation coefficient matters}.
In: \bbtitle{Proceedings of the Evaluation and Assessment in Software
  Engineering},
pp. \bfpage{120}--\blpage{129}
(\byear{2020})
\end{bchapter}
\endbibitem

\bibitem{yao2021impact}
\begin{barticle}
\bauthor{\bsnm{Yao}, \binits{J.}},
\bauthor{\bsnm{Shepperd}, \binits{M.}}:
\batitle{The impact of using biased performance metrics on software defect
  prediction research}.
\bjtitle{Information and Software Technology}
\bvolume{139},
\bfpage{106664}
(\byear{2021})
\end{barticle}
\endbibitem

\bibitem{boughorbel2017optimal}
\begin{barticle}
\bauthor{\bsnm{Boughorbel}, \binits{S.}},
\bauthor{\bsnm{Jarray}, \binits{F.}},
\bauthor{\bsnm{El-Anbari}, \binits{M.}}:
\batitle{Optimal classifier for imbalanced data using matthews correlation
  coefficient metric}.
\bjtitle{PloS one}
\bvolume{12}(\bissue{6}),
\bfpage{0177678}
(\byear{2017})
\end{barticle}
\endbibitem

\bibitem{chicco2017ten}
\begin{barticle}
\bauthor{\bsnm{Chicco}, \binits{D.}}:
\batitle{Ten quick tips for machine learning in computational biology}.
\bjtitle{BioData mining}
\bvolume{10}(\bissue{1}),
\bfpage{1}--\blpage{17}
(\byear{2017})
\end{barticle}
\endbibitem

\bibitem{zainab2020performance}
\begin{bchapter}
\bauthor{\bsnm{Zainab}, \binits{A.}},
\bauthor{\bsnm{Ghrayeb}, \binits{A.}},
\bauthor{\bsnm{Houchati}, \binits{M.}},
\bauthor{\bsnm{Refaat}, \binits{S.S.}},
\bauthor{\bsnm{Abu-Rub}, \binits{H.}}:
\bctitle{Performance evaluation of tree-based models for big data load
  forecasting using randomized hyperparameter tuning}.
In: \bbtitle{2020 IEEE International Conference on Big Data (Big Data)},
pp. \bfpage{5332}--\blpage{5339}
(\byear{2020}).
\bcomment{IEEE}
\end{bchapter}
\endbibitem

\bibitem{sandha2020mango}
\begin{bchapter}
\bauthor{\bsnm{Sandha}, \binits{S.S.}},
\bauthor{\bsnm{Aggarwal}, \binits{M.}},
\bauthor{\bsnm{Fedorov}, \binits{I.}},
\bauthor{\bsnm{Srivastava}, \binits{M.}}:
\bctitle{Mango: A python library for parallel hyperparameter tuning}.
In: \bbtitle{ICASSP 2020-2020 IEEE International Conference on Acoustics,
  Speech and Signal Processing (ICASSP)},
pp. \bfpage{3987}--\blpage{3991}
(\byear{2020}).
\bcomment{IEEE}
\end{bchapter}
\endbibitem

\bibitem{wilcoxon1992individual}
\begin{bchapter}
\bauthor{\bsnm{Wilcoxon}, \binits{F.}}:
\bctitle{Individual comparisons by ranking methods}.
In: \bbtitle{Breakthroughs in Statistics},
pp. \bfpage{196}--\blpage{202}.
\bpublisher{Springer},
\blocation{New York City, USA}
(\byear{1992})
\end{bchapter}
\endbibitem

\bibitem{pan2022test}
\begin{barticle}
\bauthor{\bsnm{Pan}, \binits{R.}},
\bauthor{\bsnm{Bagherzadeh}, \binits{M.}},
\bauthor{\bsnm{Ghaleb}, \binits{T.A.}},
\bauthor{\bsnm{Briand}, \binits{L.}}:
\batitle{Test case selection and prioritization using machine learning: a
  systematic literature review}.
\bjtitle{Empirical Software Engineering}
\bvolume{27}(\bissue{2}),
\bfpage{1}--\blpage{43}
(\byear{2022})
\end{barticle}
\endbibitem

\bibitem{wang2016empirical}
\begin{botherref}
\oauthor{\bsnm{Wang}, \binits{R.}},
\oauthor{\bsnm{Jiang}, \binits{S.}},
\oauthor{\bsnm{Chen}, \binits{D.}},
\oauthor{\bsnm{Zhang}, \binits{Y.}}:
Empirical study of the effects of different similarity measures on test case
  prioritization.
Mathematical Problems in Engineering
\textbf{2016}
(2016)
\end{botherref}
\endbibitem

\bibitem{davies1979cluster}
\begin{botherref}
\oauthor{\bsnm{Davies}, \binits{D.L.}},
\oauthor{\bsnm{Bouldin}, \binits{D.W.}}:
A cluster separation measure.
IEEE transactions on pattern analysis and machine intelligence
(2),
224--227
(1979)
\end{botherref}
\endbibitem

\bibitem{noor2017studying}
\begin{bchapter}
\bauthor{\bsnm{Noor}, \binits{T.B.}},
\bauthor{\bsnm{Hemmati}, \binits{H.}}:
\bctitle{Studying test case failure prediction for test case prioritization}.
In: \bbtitle{Proceedings of the 13th International Conference on Predictive
  Models and Data Analytics in Software Engineering},
pp. \bfpage{2}--\blpage{11}
(\byear{2017})
\end{bchapter}
\endbibitem

\bibitem{luo2018assessing}
\begin{bchapter}
\bauthor{\bsnm{Luo}, \binits{Q.}},
\bauthor{\bsnm{Moran}, \binits{K.}},
\bauthor{\bsnm{Poshyvanyk}, \binits{D.}},
\bauthor{\bsnm{Di~Penta}, \binits{M.}}:
\bctitle{Assessing test case prioritization on real faults and mutants}.
In: \bbtitle{2018 IEEE International Conference on Software Maintenance and
  Evolution (ICSME)},
pp. \bfpage{240}--\blpage{251}
(\byear{2018}).
\bcomment{IEEE}
\end{bchapter}
\endbibitem

\bibitem{luo2016large}
\begin{bchapter}
\bauthor{\bsnm{Luo}, \binits{Q.}},
\bauthor{\bsnm{Moran}, \binits{K.}},
\bauthor{\bsnm{Poshyvanyk}, \binits{D.}}:
\bctitle{A large-scale empirical comparison of static and dynamic test case
  prioritization techniques}.
In: \bbtitle{Proceedings of the 2016 24th ACM SIGSOFT International Symposium
  on Foundations of Software Engineering},
pp. \bfpage{559}--\blpage{570}
(\byear{2016})
\end{bchapter}
\endbibitem

\bibitem{kwon2014test}
\begin{bchapter}
\bauthor{\bsnm{Kwon}, \binits{J.-H.}},
\bauthor{\bsnm{Ko}, \binits{I.-Y.}},
\bauthor{\bsnm{Rothermel}, \binits{G.}},
\bauthor{\bsnm{Staats}, \binits{M.}}:
\bctitle{Test case prioritization based on information retrieval concepts}.
In: \bbtitle{2014 21st Asia-Pacific Software Engineering Conference},
vol. \bseriesno{1},
pp. \bfpage{19}--\blpage{26}
(\byear{2014}).
\bcomment{IEEE}
\end{bchapter}
\endbibitem

\bibitem{luo2018static}
\begin{barticle}
\bauthor{\bsnm{Luo}, \binits{Q.}},
\bauthor{\bsnm{Moran}, \binits{K.}},
\bauthor{\bsnm{Zhang}, \binits{L.}},
\bauthor{\bsnm{Poshyvanyk}, \binits{D.}}:
\batitle{How do static and dynamic test case prioritization techniques perform
  on modern software systems? an extensive study on github projects}.
\bjtitle{IEEE Transactions on Software Engineering}
\bvolume{45}(\bissue{11}),
\bfpage{1054}--\blpage{1080}
(\byear{2018})
\end{barticle}
\endbibitem

\bibitem{lachmann2016system}
\begin{bchapter}
\bauthor{\bsnm{Lachmann}, \binits{R.}},
\bauthor{\bsnm{Schulze}, \binits{S.}},
\bauthor{\bsnm{Nieke}, \binits{M.}},
\bauthor{\bsnm{Seidl}, \binits{C.}},
\bauthor{\bsnm{Schaefer}, \binits{I.}}:
\bctitle{System-level test case prioritization using machine learning}.
In: \bbtitle{2016 15th IEEE International Conference on Machine Learning and
  Applications (ICMLA)},
pp. \bfpage{361}--\blpage{368}
(\byear{2016}).
\bcomment{IEEE}
\end{bchapter}
\endbibitem

\bibitem{arafeen2013test}
\begin{bchapter}
\bauthor{\bsnm{Arafeen}, \binits{M.J.}},
\bauthor{\bsnm{Do}, \binits{H.}}:
\bctitle{Test case prioritization using requirements-based clustering}.
In: \bbtitle{2013 IEEE Sixth International Conference on Software Testing,
  Verification and Validation},
pp. \bfpage{312}--\blpage{321}
(\byear{2013}).
\bcomment{IEEE}
\end{bchapter}
\endbibitem

\bibitem{laali2016test}
\begin{bchapter}
\bauthor{\bsnm{Laali}, \binits{M.}},
\bauthor{\bsnm{Liu}, \binits{H.}},
\bauthor{\bsnm{Hamilton}, \binits{M.}},
\bauthor{\bsnm{Spichkova}, \binits{M.}},
\bauthor{\bsnm{Schmidt}, \binits{H.W.}}:
\bctitle{Test case prioritization using online fault detection information}.
In: \bbtitle{Ada-Europe International Conference on Reliable Software
  Technologies},
pp. \bfpage{78}--\blpage{93}
(\byear{2016}).
\bcomment{Springer}
\end{bchapter}
\endbibitem

\bibitem{engstrom2010empirical}
\begin{bchapter}
\bauthor{\bsnm{Engstr{\"o}m}, \binits{E.}},
\bauthor{\bsnm{Runeson}, \binits{P.}},
\bauthor{\bsnm{Wikstrand}, \binits{G.}}:
\bctitle{An empirical evaluation of regression testing based on fix-cache
  recommendations}.
In: \bbtitle{2010 Third International Conference on Software Testing,
  Verification and Validation},
pp. \bfpage{75}--\blpage{78}
(\byear{2010}).
\bcomment{IEEE}
\end{bchapter}
\endbibitem

\bibitem{anderson2014improving}
\begin{bchapter}
\bauthor{\bsnm{Anderson}, \binits{J.}},
\bauthor{\bsnm{Salem}, \binits{S.}},
\bauthor{\bsnm{Do}, \binits{H.}}:
\bctitle{Improving the effectiveness of test suite through mining historical
  data}.
In: \bbtitle{Proceedings of the 11th Working Conference on Mining Software
  Repositories},
pp. \bfpage{142}--\blpage{151}
(\byear{2014}).
\bcomment{ACM}
\end{bchapter}
\endbibitem

\bibitem{kim2010effective}
\begin{bchapter}
\bauthor{\bsnm{Kim}, \binits{S.}},
\bauthor{\bsnm{Baik}, \binits{J.}}:
\bctitle{An effective fault aware test case prioritization by incorporating a
  fault localization technique}.
In: \bbtitle{Proceedings of the 2010 ACM-IEEE International Symposium on
  Empirical Software Engineering and Measurement},
p. \bfpage{5}
(\byear{2010}).
\bcomment{ACM}
\end{bchapter}
\endbibitem

\bibitem{wang2016enhancing}
\begin{bchapter}
\bauthor{\bsnm{Wang}, \binits{S.}},
\bauthor{\bsnm{Ali}, \binits{S.}},
\bauthor{\bsnm{Yue}, \binits{T.}},
\bauthor{\bsnm{Bakkeli}, \binits{{\O}.}},
\bauthor{\bsnm{Liaaen}, \binits{M.}}:
\bctitle{Enhancing test case prioritization in an industrial setting with
  resource awareness and multi-objective search}.
In: \bbtitle{Proceedings of the 38th International Conference on Software
  Engineering Companion},
pp. \bfpage{182}--\blpage{191}
(\byear{2016})
\end{bchapter}
\endbibitem

\bibitem{mondal2015exploring}
\begin{bchapter}
\bauthor{\bsnm{Mondal}, \binits{D.}},
\bauthor{\bsnm{Hemmati}, \binits{H.}},
\bauthor{\bsnm{Durocher}, \binits{S.}}:
\bctitle{Exploring test suite diversification and code coverage in
  multi-objective test case selection}.
In: \bbtitle{2015 IEEE 8th International Conference on Software Testing,
  Verification and Validation (ICST)},
pp. \bfpage{1}--\blpage{10}
(\byear{2015}).
\bcomment{IEEE}
\end{bchapter}
\endbibitem

\bibitem{pradhan2019employing}
\begin{barticle}
\bauthor{\bsnm{Pradhan}, \binits{D.}},
\bauthor{\bsnm{Wang}, \binits{S.}},
\bauthor{\bsnm{Ali}, \binits{S.}},
\bauthor{\bsnm{Yue}, \binits{T.}},
\bauthor{\bsnm{Liaaen}, \binits{M.}}:
\batitle{Employing rule mining and multi-objective search for dynamic test case
  prioritization}.
\bjtitle{Journal of Systems and Software}
\bvolume{153},
\bfpage{86}--\blpage{104}
(\byear{2019})
\end{barticle}
\endbibitem

\bibitem{miranda2018fast}
\begin{bchapter}
\bauthor{\bsnm{Miranda}, \binits{B.}},
\bauthor{\bsnm{Cruciani}, \binits{E.}},
\bauthor{\bsnm{Verdecchia}, \binits{R.}},
\bauthor{\bsnm{Bertolino}, \binits{A.}}:
\bctitle{Fast approaches to scalable similarity-based test case
  prioritization}.
In: \bbtitle{2018 IEEE/ACM 40th International Conference on Software
  Engineering (ICSE)},
pp. \bfpage{222}--\blpage{232}
(\byear{2018}).
\bcomment{IEEE}
\end{bchapter}
\endbibitem

\bibitem{spieker2017reinforcement}
\begin{bchapter}
\bauthor{\bsnm{Spieker}, \binits{H.}},
\bauthor{\bsnm{Gotlieb}, \binits{A.}},
\bauthor{\bsnm{Marijan}, \binits{D.}},
\bauthor{\bsnm{Mossige}, \binits{M.}}:
\bctitle{Reinforcement learning for automatic test case prioritization and
  selection in continuous integration}.
In: \bbtitle{Proceedings of the 26th ACM SIGSOFT International Symposium on
  Software Testing and Analysis},
pp. \bfpage{12}--\blpage{22}
(\byear{2017})
\end{bchapter}
\endbibitem

\bibitem{bagherzadeh2021reinforcement}
\begin{botherref}
\oauthor{\bsnm{Bagherzadeh}, \binits{M.}},
\oauthor{\bsnm{Kahani}, \binits{N.}},
\oauthor{\bsnm{Briand}, \binits{L.}}:
Reinforcement learning for test case prioritization.
IEEE Transactions on Software Engineering
(2021)
\end{botherref}
\endbibitem

\bibitem{tan2016introduction}
\begin{bbook}
\bauthor{\bsnm{Tan}, \binits{P.-N.}},
\bauthor{\bsnm{Steinbach}, \binits{M.}},
\bauthor{\bsnm{Kumar}, \binits{V.}}:
\bbtitle{Introduction to Data Mining}.
\bpublisher{Pearson Education},
\blocation{Noida, Plot–C01, Sector-16}
(\byear{2016})
\end{bbook}
\endbibitem

\bibitem{chen2016similarity}
\begin{barticle}
\bauthor{\bsnm{Chen}, \binits{J.}},
\bauthor{\bsnm{Kuo}, \binits{F.-C.}},
\bauthor{\bsnm{Chen}, \binits{T.Y.}},
\bauthor{\bsnm{Towey}, \binits{D.}},
\bauthor{\bsnm{Su}, \binits{C.}},
\bauthor{\bsnm{Huang}, \binits{R.}}:
\batitle{A similarity metric for the inputs of oo programs and its application
  in adaptive random testing}.
\bjtitle{IEEE Transactions on Reliability}
\bvolume{66}(\bissue{2}),
\bfpage{373}--\blpage{402}
(\byear{2016})
\end{barticle}
\endbibitem

\bibitem{zhao2015clustering}
\begin{bchapter}
\bauthor{\bsnm{Zhao}, \binits{X.}},
\bauthor{\bsnm{Wang}, \binits{Z.}},
\bauthor{\bsnm{Fan}, \binits{X.}},
\bauthor{\bsnm{Wang}, \binits{Z.}}:
\bctitle{A clustering-bayesian network based approach for test case
  prioritization}.
In: \bbtitle{2015 IEEE 39th Annual Computer Software and Applications
  Conference},
vol. \bseriesno{3},
pp. \bfpage{542}--\blpage{547}
(\byear{2015}).
\bcomment{IEEE}
\end{bchapter}
\endbibitem

\bibitem{mirarab2007prioritization}
\begin{bchapter}
\bauthor{\bsnm{Mirarab}, \binits{S.}},
\bauthor{\bsnm{Tahvildari}, \binits{L.}}:
\bctitle{A prioritization approach for software test cases based on bayesian
  networks}.
In: \bbtitle{International Conference on Fundamental Approaches to Software
  Engineering},
pp. \bfpage{276}--\blpage{290}
(\byear{2007}).
\bcomment{Springer}
\end{bchapter}
\endbibitem

\bibitem{hasan2017test}
\begin{bchapter}
\bauthor{\bsnm{Hasan}, \binits{M.A.}},
\bauthor{\bsnm{Rahman}, \binits{M.A.}},
\bauthor{\bsnm{Siddik}, \binits{M.S.}}:
\bctitle{Test case prioritization based on dissimilarity clustering using
  historical data analysis}.
In: \bbtitle{International Conference on Information, Communication and
  Computing Technology},
pp. \bfpage{269}--\blpage{281}
(\byear{2017}).
\bcomment{Springer}
\end{bchapter}
\endbibitem

\bibitem{pang2013identifying}
\begin{bchapter}
\bauthor{\bsnm{Pang}, \binits{Y.}},
\bauthor{\bsnm{Xue}, \binits{X.}},
\bauthor{\bsnm{Namin}, \binits{A.S.}}:
\bctitle{Identifying effective test cases through k-means clustering for
  enhancing regression testing}.
In: \bbtitle{2013 12th International Conference on Machine Learning and
  Applications},
vol. \bseriesno{2},
pp. \bfpage{78}--\blpage{83}
(\byear{2013}).
\bcomment{IEEE}
\end{bchapter}
\endbibitem

\end{thebibliography}


\end{document}